\documentclass[12pt,preprint]{emulateapj}
\usepackage{amsmath,natbib,graphicx}

\usepackage{amssymb}
\submitted{The Astrophysical Journal, accepted}
\usepackage[caption=false]{subfig}
\usepackage{epsf}
\usepackage{color}
\usepackage{afterpage}
\input{epsf}
\usepackage{epsfig}
\DeclareGraphicsExtensions{.jpg,.pdf,.png,.eps,.ps}
\newcommand{\ltsima}{$\; \buildrel < \over \sim \;$}
\newcommand{\ltsim}{\lower.5ex\hbox{\ltsima}}

\def\Munich{1}
\def\ExcellenceCluster{2}
\def\Michigan{3}
\def\Illinois{4}
\def\NCSA{5}
\def\MPE{6}
\def\CUTaiwan{7}
\def\UNDakota{8}
\def\Fermilab{9}

\def\IAPFrance{12}
\def\CfA{13}
\def\KECKfellow{14}
\def\UCOLick{15}
\def\UChicago{16}
\def\Tokyo{17}
\def\IAAASTaiwan{18}
\def\Harvard{19}
\def\STSI{20}
\def\NOAO{21}

\begin{document}
\title{A Multiband Study of the  Galaxy Populations of the First Four \\
Sunyaev--Zeldovich Effect selected Galaxy Clusters}

\author{A. Zenteno,\altaffilmark{\Munich,\ExcellenceCluster}
J. Song,\altaffilmark{\Michigan}
S. Desai,\altaffilmark{\Illinois,\NCSA}
R. Armstrong,\altaffilmark{\NCSA}
J. J. Mohr,\altaffilmark{\Munich,\ExcellenceCluster,\MPE} 
C.-C. Ngeow,\altaffilmark{\CUTaiwan,\Illinois}
W. A. Barkhouse,\altaffilmark{\UNDakota}\\
S. S. Allam,\altaffilmark{\Fermilab}
K. Andersson,\altaffilmark{\Munich}
G. Bazin,\altaffilmark{\Munich,\ExcellenceCluster}
B. A. Benson,\altaffilmark{\UChicago} 
E. Bertin,\altaffilmark{\IAPFrance} 
M. Brodwin, \altaffilmark{\CfA,\KECKfellow}\\
E. J. Buckley-Geer,\altaffilmark{\Fermilab}
S. M. Hansen,\altaffilmark{\UCOLick}
F. W. High,\altaffilmark{\UChicago}
H. Lin,\altaffilmark{\Fermilab}
Y.-T. Lin,\altaffilmark{\Tokyo,\IAAASTaiwan}
J. Liu,\altaffilmark{\Munich,\ExcellenceCluster,\UChicago}
A. Rest,\altaffilmark{\STSI}\\
R. C. Smith,\altaffilmark{\NOAO}
B. Stalder,\altaffilmark{\Harvard} 
A. A. Stark,\altaffilmark{\CfA} 
D. L. Tucker,\altaffilmark{\Fermilab} and
Y. Yang\altaffilmark{\Illinois}
}

\altaffiltext{\Munich}{Department of Physics, Ludwig-Maximilians-Universit\"{a}t, Scheinerstr.\ 1, 81679 M\"{u}nchen, Germany}
\altaffiltext{\ExcellenceCluster}{Excellence Cluster Universe, Boltzmannstr.\ 2, 85748 Garching, Germany}
\altaffiltext{\Michigan}{Department of Physics, University of Michigan, 450 Church St. Ann Arbor, MI 48109}
\altaffiltext{\Illinois}{Department of Astronomy, University of Illinois, 1002 West Green Street, Urbana, IL 61801}
\altaffiltext{\NCSA}{National Center for Supercomputing Applications, University of Illinois, 1205 West Clark Street, Urbanan, IL 61801}
\altaffiltext{\MPE}{Max-Planck-Institut f\"{u}r extraterrestrische Physik, Giessenbachstr.\ 85748 Garching, Germany}
\altaffiltext{\CUTaiwan}{Graduate Institute of Astronomy, National Central University, No. 300 Jonghda Rd, Jhongli City 32001 Taiwan}
\altaffiltext{\UNDakota}{Department of Physics \& Astrophysics, University of North Dakota, Grand Forks, ND 58202}
\altaffiltext{\Fermilab}{Fermi National Accelerator Laboratory, P.O. Box 500, Batavia, IL 60510}
\altaffiltext{\IAPFrance}{Institut d'Astrophysique de Paris, UMR 7095 CNRS, Universit\'e Pierre et Marie Curie, 98 bis boulevard Arago, F-75014 Paris, France}
\altaffiltext{\CfA}{Harvard-Smithsonian Center for Astrophysics, 60 Garden Street, Cambridge, MA 02138}
\altaffiltext{\KECKfellow}{W. M. Keck Postdoctoral Fellow at the Harvard-Smithsonian Center for Astrophysics}
\altaffiltext{\UCOLick}{University of California Observatories \& Department of Astronomy, University of California, Santa Cruz, CA 95064}
\altaffiltext{\UChicago}{University of Chicago, 5640 South Ellis Avenue, Chicago, IL 60637}
\altaffiltext{\Tokyo}{Institute for Physics and Mathematics of the Universe, University of Tokyo, 5-1-5 Kashiwa-no-ha, Kashiwa-shi, Chiba 277- 8568, Japan}
\altaffiltext{\IAAASTaiwan}{Institute of Astronomy \& Astrophysics, Academia Sinica, Taipei, Taiwan}
\altaffiltext{\Harvard}{Department of Physics, Harvard University, 17 Oxford Street, Cambridge, MA 02138}
\altaffiltext{\STSI}{Space Telescope Science Institute, 3700 San Martin Dr., Baltimore, MD 21218}
\altaffiltext{\NOAO}{Cerro Tololo Inter-American Observatory, National Optical Astronomy Observatory, La Serena, Chile}

\email{alfredo@usm.lmu.de}

\begin{abstract}

We present first  results of an examination of  the optical properties
of  the galaxy  populations in  SZE selected  galaxy  clusters.  Using
clusters  selected  by  the  South  Pole  Telescope  survey  and  deep
multiband optical  data from the  Blanco Cosmology Survey,  we measure
the radial profile, the luminosity function, the blue fraction and the
halo  occupation  number  of  the  galaxy populations  of  these  four
clusters  with  redshifts ranging  from  0.3 to  1.   Our  goal is  to
understand whether there are  differences among the galaxy populations
of  these  SZE  selected  clusters  and  previously  studied  clusters
selected in  the optical and  the X-ray.  The radial  distributions of
galaxies in the  four systems are consistent with  NFW profiles with a
galaxy  concentration of  3 to  6.   We show  that the  characteristic
luminosities in  $griz$ bands  are consistent with  passively evolving
populations emerging from a single burst at redshift $z=3$.  The faint
end  power law  slope of  the luminosity  function is  found to  be on
average $\alpha \approx -1.2$  in $griz$.  Halo occupation numbers (to
$m^*+2$) for these systems appear to be consistent with those based on
X-ray selected  clusters.  The  blue fraction estimated  to $0.36L^*$,
for  the  three lower  redshift  systems,  suggests  an increase  with
redshift, although with the current sample the uncertainties are still
large.  Overall, this pilot study  of the first four clusters provides
no  evidence  that the  galaxy  populations  in  these systems  differ
significantly  from those  in previously  studied  cluster populations
selected in the X-ray or the optical.
\end{abstract}


\bigskip\bigskip

\section{Introduction}
\label{sec:intro}

Galaxy  clusters  can be  readily discovered or selected using optical or IR emission from their member galaxies, X-ray emission from the hot intracluster medium and now even by the impact of this intracluster medium on the cosmic microwave background temperature toward these systems.  First,  from  optical   observations,  \citet{abell58}
identified,  catalogued and characterized  clusters of  galaxies using
classification   criteria    like   $compactness$,   $distance$,   and
$richness$.  Later, new optical surveys added other optical properties
to the clusters.  Luminosity  function, radial profile, blue fraction,
dwarf-to-giant  ratio, among  others, became  tools  for understanding
different physical processes in the galaxy cluster environment.

With the advent of space based astronomy new properties of clusters of
galaxies  were  discovered.  Strong  X-ray  emission  made the  galaxy
clusters some of the most  luminous objects in the universe, and their
properties  like X-ray  luminosity,  temperature, and  mass have  been
compiled  in several X-ray  selected cluster  surveys \citep[see,][for
example]{giacconi72,voges99,voges00,bohringer04}.

In the infra-red regime, the  properties of clusters have been studied
mainly  relying  on  the   X-ray  or  optical  cluster  identification
\citep[see,][                                                     among
  others]{depropris99,lin03b,lin04a,toft04,depropris07,muzzin07a,muzzin07b,roncarelli10}.
From  IR selected  clusters, some  of the  first studies  analyzed the
cluster      populations     based     on      individual     clusters
\citep{stanford97,stanford05}.  Later, systematic searches of clusters
in the infrared became feasible with the operation of space telescopes
and with ground based  telescopes with advanced IR detectors.  Surveys
such  as FLAMEX  \citep{elston06}, UKIDSS  \citep{vanbreukelen06}, FLS
\citep{muzzin08} and the IRAC Shallow Survey \citep{eisenhardt08} have
delivered  cluster  catalogs,   at  high  redshift,  allowing  initial
systematic  characterization  of   the  galaxy  populations  on  those
systems.

In  the millimeter regime,  the use  of the  Sunyaev-Zel'dovich Effect
\citep[SZE,][]{sunyaev72}  as  a   selection  method  for  cluster  of
galaxies     has    recently     produced     the    first     results
\citep{staniszewski09,vanderlinde10}.   The  use  of the  SZE
effect for cluster detection has several advantages.  A catalog of SZE
selected  clusters  is  approximately  mass limited,  nearly  redshift
independent  and the observable  signature is  closely related  to the
cluster mass \citep{birkinshaw99,carlstrom02}, making it less prone to
be biased  in the selection.   In particular, an SZE  selected cluster
sample  provides an  opportunity  to systematically  study the  galaxy
populations and  its redshift  evolution in clusters  of the  same mass
range over a wide range of redshift.

In this paper we use tools developed for optical studies to analyze the galaxy populations of the first four SZE selected clusters published by the  South Pole  Telescope (SPT) collaboration \citep{staniszewski09}.   As well as being among the first SZ selected systems, these clusters are among the most well studied.  This sample has been imaged deeply in the optical Blanco Cosmology Survey, studied in the X-ray \citep{andersson10}, targeted spectroscopically for redshifts \citep{high10}, and the BCS data have been used to estimate weak lensing masses \citep{mcinnes09}.  Also these four systems span a broad range in redshift and mass, much like the larger samples that have been published so far \citep{vanderlinde10, williamson11}.  In this pilot study, we study the luminosity function, the radial profile, the Halo Occupation number and the blue fraction, in an effort to answer a basic question: Are the galaxy populations from these first SZE  selected clusters any different than the populations in clusters selected by other means?

The paper is organized as  follows: \S2 describes the observations and
data reduction.  In section \S3,  properties of the clusters,  such as
redshift  and  mass,  are  described.   In \S4  we  study  the  galaxy
populations in the clusters,  presenting the main results.  Conclusion
of this study are presented  at section \S5.  Magnitudes are quoted in
AB system.

We  assume a  flat,  $\Lambda$CDM  cosmology with  $H_0$  = 100$h$  km
s$^{-1}$  M$pc^{-1}$, $h$  =  0.702, and  matter  density $\Omega_m$  =
0.272, according to WMAP7 + BAO + $H_0$ data \citep{komatsu10}.

\section{Observations and Data Reduction}
\subsection{Blanco Cosmology Survey}
\label{sec:observations}

The          Blanco          Cosmology          Survey\footnote[1]{\tt
  http://cosmology.illinois.edu/BCS/}  (BCS)  project  was awarded  60
nights from  the NOAO (National Optical  Astronomy Observatory) survey
program starting in semester  2005B.  Data were gathered in 2005--2008
using   the  Blanco   4-meter  telescope   located  at   Cerro  Tololo
Inter-American Observatory\footnote[2]{\tt Cerro Tololo Inter-American
  Observatory  (CTIO)  is a  division  of  the  U.S. National  Optical
  Astronomy Observatory  (NOAO), which is operated  by the Association
  of  Universities for  Research in  Astronomy (AURA),  under contract
  with the  National Science Foundation.} in Chile.   The telescope is
equipped with  a wide  field camera called  the Mosaic2  imager, which
consists of an array of eight  $2K\times 4K$ CCDs.  The pixel scale of
Mosaic2 imager is  $0.27$ arc-second per pixel, leading  to a field of
view of about $0.36$ square degree.  The observations were carried out
to obtain a deep, four band photometric survey ($g$, $r$, $i$ and $z$)
of  two  $50$  deg$^2$  patches   of  the  southern  sky  centered  at
23h00m,-55$^{\circ}$12'' and  05h30m,-55$^{\circ}$47''.  These regions
were  chosen  to enable  observations  by  three mm-wavelength  survey
experiments (the SPT, the Atacama Pathfinder Experiment (APEX) and the
Atacama Cosmology Telescope (ACT) experiments).  On photometric nights
we also observed several standard  star fields that contain stars with
known magnitudes.  This approach allows the calibration of our data to
the standard magnitude system.   In addition, we obtained deep imaging
of  several  fields  overlapping  published spectroscopic  surveys  to
enable  calibration of photo-z's  using samples  of many  thousands of
spectroscopic  redshifts.  The data  volume we  collected for  the BCS
observation was about $20$ to $30$ Gigabytes/night.

The first  three seasons (2005 to  2007) of the BCS  imaging data were
processed  in 2008  and 2009  using version  3 of the data  management system
developed for  the upcoming Dark  Energy Survey (DES). Details  of the
DES  data  management  system  can  be found  in  \citet{ngeow06}  and
\citet{mohr08}. A  brief description is presented in  this paper. Data
parallel processing was carried out primarily on NCSA's TeraGrid IA-64
Linux cluster. The pipeline  processing middleware developed within
the  DES data management  system provides  the infrastructure  for the
automated and robust execution of our parallel pipeline processing on
the TeraGrid cluster.

To  remove  the  instrumental  signatures,  the raw  BCS  images  were
processed  using  the  following  corrections:  crosstalk  correction,
overscan  correction,  bias  subtraction,  flat fielding,  fringe  and
illumination correction.  Bad columns and pixels, saturated pixels and
bright star  halos, and bleed  trails are masked  automatically.  Wide
field   imagers  have   field  distortions   that   generally  deviate
significantly  from a simple  tangent plane,  and there  are typically
telescope pointing  errors as well.   The AstrOmatic code  {\tt SCAMP}
\citep{bertin06}  was  used  to  refine the  astrometric  solution  by
matching the detected  stars in BCS images to  the USNO-B catalog.  We
adopted the  $PV$ distortion model  that maps detector  coordinates to
sky   coordinate  using   a  third   order  polynomial   expansion  of
distortions, across  each CCD, relative  to a tangent plane.   The DES
data  management  system  is  using  an experimental  version  of  the
AstrOmatic tool {\tt  SExtractor} \citep{bertin96}.  This experimental
version  includes  model  fitting  photometry and  improved  modes  of
star-galaxy classification to  detect and catalog astronomical objects
in  the  images.   We  harvested  a  wide  range  of  photometric  and
astrometric  measurements (and  their uncertainties)  for  each object
during this cataloging.

For the  photometric nights that  include observation of  the standard
star fields, we determined the band dependent (atmospheric) extinction
coefficients ($k$)  together with  CCD and band  dependent photometric
zeropoints ($a$) and instrumental color terms ($b$). Specifically, the
equation we constructed  for each star in the  standard star fields is
$m_{inst} -  m_{std} = \sum_i w_i\times  [a_i + b_i(\Delta  C)] + kX$,
where  $w_i =  1$  if the  standard  star is  on CCD  $i$;  $w_i =  0$
otherwise.   In  this  equation,  $m_{inst}$  and  $m_{std}$  are  the
instrumental  and   the  true  magnitudes  for   the  standard  stars,
respectively, $\Delta  C$ is  the color offset  of the  standard stars
from a  reference color,  and $X$ is  the airmass.  The  standard star
fields include  the SDSS  Stripe 82 fields  and the  Southern Standard
Stars  Network fields\footnote[3]{{\tt
    http://www-star.fnal.gov/Southern\_ugriz/index.html}}.          The
resulting  photometric  solutions  were  then used  to  calibrate  the
magnitudes for other astronomical objects observed on the same night.

The nightly  reduced and astrometric refined images  were remapped and
coadded to a pre-defined grid of tiles (which is a rectangular tangent
plane projection,  with $\sim36$ arc-minute  on a side  (hereafter the
BCS  tiles) in  the sky  using  another AstrOmatic  tool, {\tt  SWarp}
\citep{bertin02}.   During   this  coaddition  we  carry   out  a  PSF
homogenization across each tile and  within each band to match the PSF
to median  delivered seeing in that  part of the  sky.  The zeropoints
for the flux scales for  these input remap images are determined using
different  sources   of  photometric  information,   including  direct
photometric zeropoints which are derived from the photometric solution
on  photometric  nights,  relative photometric  zeropoints  determined
using  all pairs  of images  that  overlap on  the sky  and the  color
behavior  of  the  stellar  locus \citep{high09}.   We  determine  the
zeropoints for all images by  doing a least squares solution using the
constraints described  above.  During  co-addition, we use  a weighted
mean combine option  in {\tt SWARP}.  The coadded  images are built in
each   band  for   a  given   coadd  tile,   then  a   $\chi^2$  image
\citep{szalay99}  is created  for detection  and cataloging  to ensure
each object will have measurements  in the $griz$ bands extracted from
the same portion of the object.

\subsection{Completeness}
\label{sec:completeness}
\begin{figure}
\begin{center}
\includegraphics[width=0.49\textwidth]{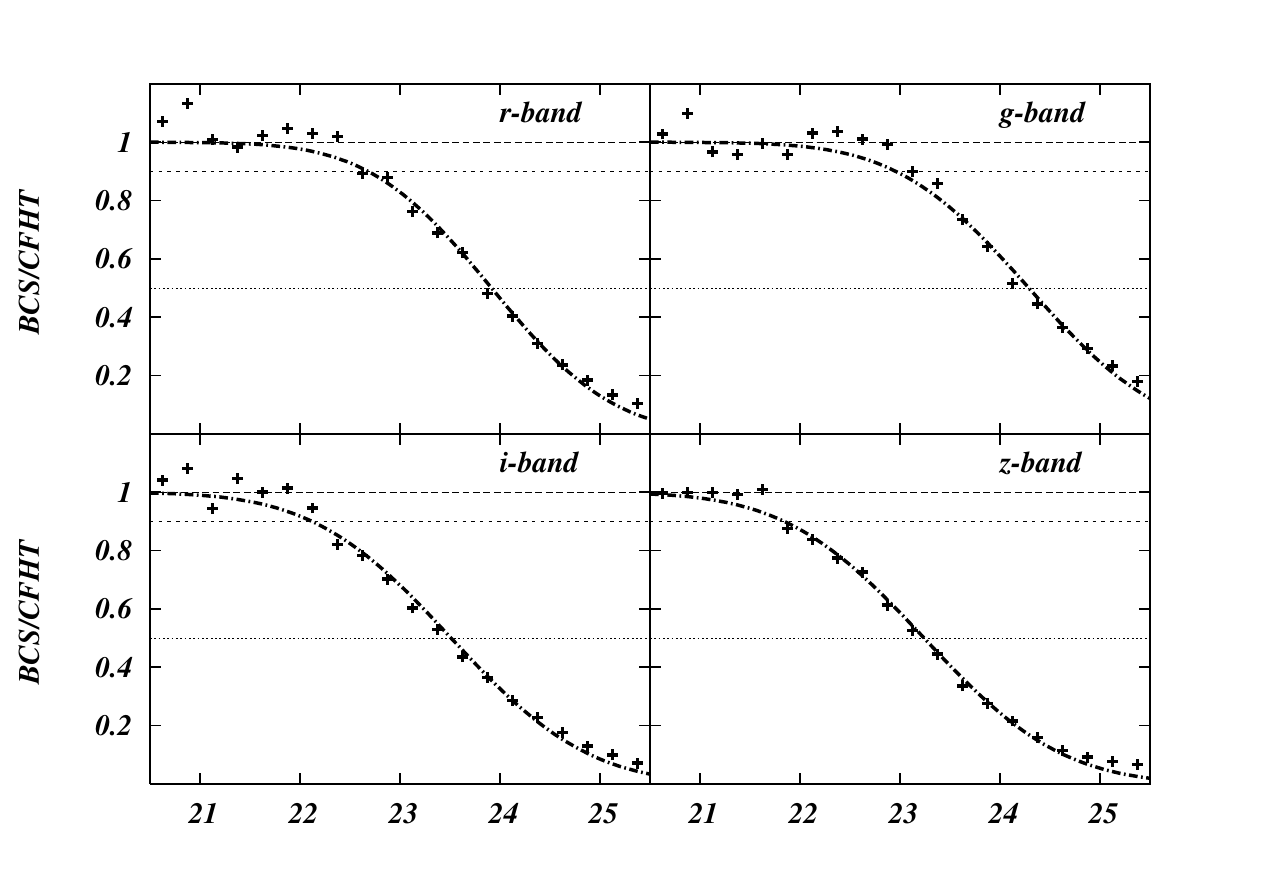}
\caption{We estimate  the completeness of our BCS  coadds by comparing
  objects  counts from  them with  counts from  deeper CFHT  data. The
  resulting completeness  curve is fitted by an  error function, which
  later is used to correct for the missing galaxies and to define 90\%
  and 50\%  completeness limits for  analysis. Here is an  example for
  the SPT-CL J0516-5430 field. \label{comp0}}
\end{center}
\end{figure}

For this work  we estimate the completeness of the  BCS tiles from the
comparison of  their $griz$ source count histograms  and those
extracted from the deeper Canada-France-Hawaii-Telescope Legacy Survey
 survey  \citep[CFHTLS,][private  communication]{brimioulle08}.  
Specifically, we used count histograms from the D-1 1 sqr.
degree  patch  at high  galactic  latitude  (l=  $172.0^{\circ}$; b  =
$-58.0^{\circ}$)  from  the  CFHTLS  Deep  Field,  whose
magnitude limit is beyond r=27 and the seeing is better than 1.0'' and
0.9'' for $g$ and  $riz$, respectively \footnote[4]{\tt Details can be
  found at  http://www.ast.obs-mip.fr/article212.html}.  Dividing both
count histograms  (see Fig.~\ref{comp0}) we can estimate  the level of
completeness in  the different  tiles in each  band.  We can  use this
completeness estimate for each field to  account for  the missing  objects  as we
approach      the     full      depth      of     the      photometry.
Table~\ref{tab:completeness}  contains the  magnitude  limits in  each
band corresponding to 50\% and 90\% completeness for the tiles used in
our analysis.

\begin{table*}
\begin{center}
\caption{Completeness  limits  for  each  tile  for  each  filter  for
  90\%/50\% completeness} \small
\begin{tabular}{ccccccc}
\hline\hline \rule[-2mm]{0mm}{6mm} 
ID & R.A. & decl. & g & r & i & z \\ 
       &[deg] & [deg] & 500 sec & 600 sec & 1350 sec  & 705 sec\\ 
\hline
SPT-CL J0516-5430 & 79.15569 & -54.50062 & 23.18/24.24 & 22.73/23.87 & 22.20/23.47 & 21.87/23.19 \\
SPT-CL J0509-5342 & 77.33908 & -53.70351 & 23.72/24.78 & 23.29/24.51 & 23.10/24.23 & 22.45/23.78 \\
SPT-CL J0528-5300 & 82.02212 & -52.99818 & 23.70/24.62 & 23.42/24.32 & 22.93/23.94 & 22.23/23.37 \\
SPT-CL J0546-5345 & 86.65700 & -53.75861 & 23.34/24.31 & 22.87/23.90 & 22.48/23.64 & 21.97/23.08 \\
\hline
\end{tabular}
\label{tab:completeness}
\end{center}
\end{table*}

\section{Basic Properties of these SPT Clusters}
\label{sec:clusters}

\begin{figure}
\begin{center}
\includegraphics[width=0.49\textwidth]{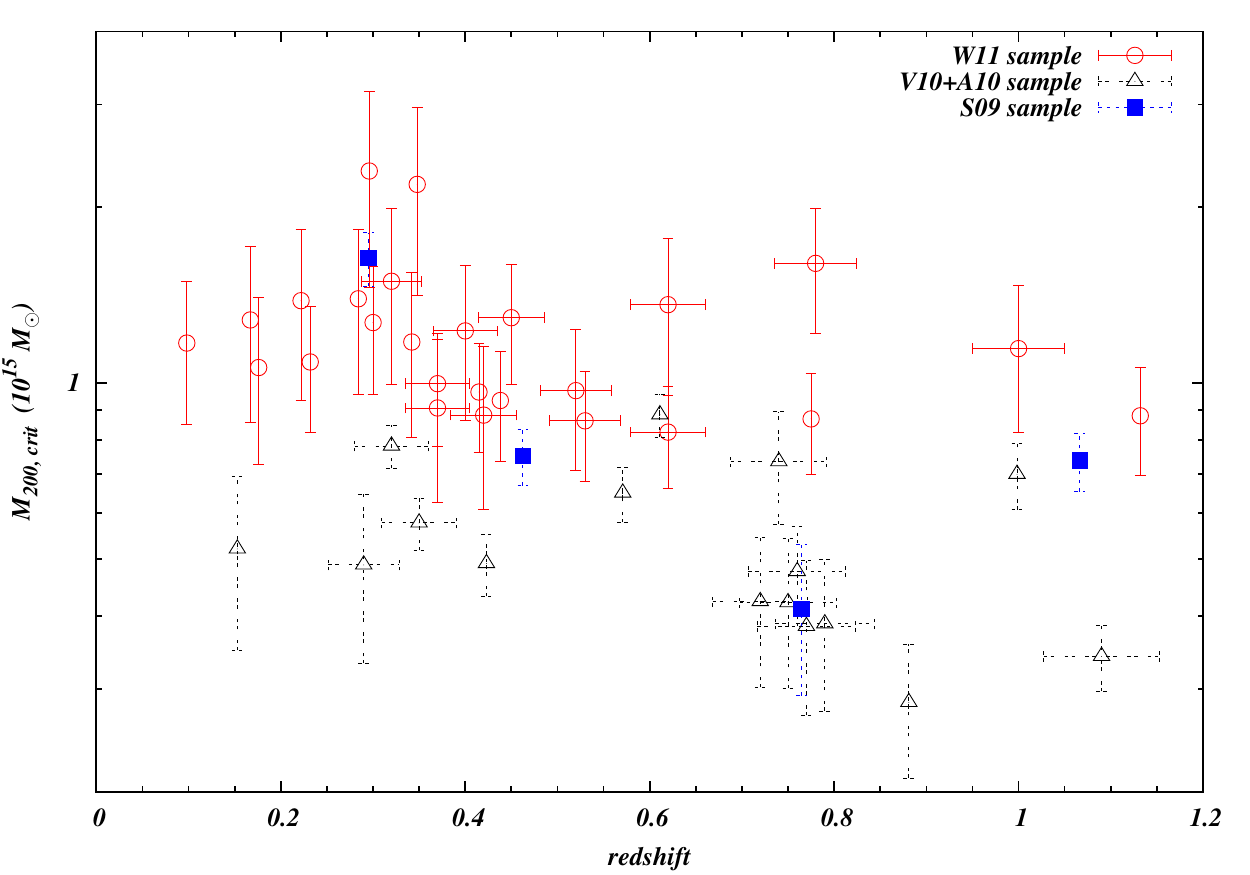}
\caption{Masses and  redshifts of the SPT cluster  sample published to
  date.   Open circles are  from \protect\citet{williamson11},  open triangles
  are from  \protect\citet{andersson10} and \protect\citet{vanderlinde10},  and filled
  squares from  \protect\citet{staniszewski09}, the sample  here studied.  The
  $M_{200,crit}$ mass estimations  come from X--ray observations where
  that is  possible, or from  the SPT detection significance.   In the
  latter case, masses have been converted from their native estimation
  $M_{200,mean}$ to $M_{200,crit}$  assuming a concentration parameter
  of $c=5$ for the halo mass (see table~\ref{tab:xclustermass}) under
  assumed  cosmology.  It can  be seen  that our  sample spans  on the
  redshift and mass space for the latest SPT sample.\label{mass_z}}
\end{center}
\end{figure}

The  basic properties of  these SPT  selected clusters,  including the
characteristics   of  the  optical   counterparts  are   presented  in
\citet{staniszewski09}  and are further  discussed in  follow on papers
\citep{menanteau09,mcinnes09,high10,andersson10}.               Several
spectroscopic  redshifts are now  available as  well as  Chandra X-ray
observations, providing  dramatically improved mass  information which
enables  the  kind  of  galaxy  population study  we  undertake  here.
 Despite being a small sample, these four clusters are
  among the most  well studied SZ selected systems  and their mass and
  redshift distributions range are  similar to the whole SPT published
  cluster  sample  (see  Fig.~\ref{mass_z}).   In  particular,  these
masses and redshifts are used to estimate the projected cluster virial
radius in which the optical properties are measured.  These properties
are presented below.

\subsection{Redshifts}
\label{sec:redshifts}

The  spectroscopic redshifts  of the  four cluster  are  now available (Table~\ref{tab:xclustermass}).
SPT-CL   J0516-5430,  a   known  cluster   identified  in   the  Abell
supplementary southern  catalog \citep[AS0520,][]{abell89} and  in the
REFLEX survey \citep[RXC J0516.6−5430,][]{bohringer04}, had a redshift
of  0.294   \citep{guzzo99}  and  0.2952   \citep{bohringer04},  values
obtained using 8 galaxy spectra from the ESO Key Programme.

For  the  other three  clusters, spectroscopic data has  recently   been
acquired.   Using LDSS-3  on Magellan  Clay  telescope, \citet{high10}
reported redshifts of  0.7648 for SPT-CL 0528-5300 and 0.4626
for SPT-CL 0509-5342.  For SPT-CL J0546-5345, \citet{brodwin10}
 used  IMACS on  Baade  Magellan telescope
to measure  a   redshift  of  1.0665. 
\newpage
\subsection{Cluster masses}
\label{sec:xmass}

As   mentioned  in  \S\ref{sec:clusters}, the  optical analyses
performed in  this work require  an estimate of the  projected cluster
virial radius.  For this  purpose, along with spectroscopic redshifts,
X-ray masses estimations are used.  We adopt mass estimates defined with respect to the critical density.

As it has been previously  mentioned, SPT-CL 0516-5430 is a previously
known cluster, and its mass has also been estimated.  With the name of
RXC J0516.6-5430 in the REFLEX survey, \citet{zhang06} used XMM-Newton
to  find a  $M_{500}$ of  (6.4 $\pm$  2.1) 10$^{14}\  M_\odot$.  Also,
recent X-ray observations of the four clusters have been performed, and
the mass estimation of SPT-CL 0517-5430 has been refined.

Using {\it Chandra} and {\it XMM-Newton}, \citet{andersson10} reported
X-ray measurements  of 15  of the 21  SZE selected clusters  presented in
\citet{vanderlinde10}.   The  observations  of  those  clusters,  which
include the  original first four clusters, have been designed  to deliver
around 1500 photons within  $0.5r_{500}$, in order to enable measurement
of  the ICM mass and ICM temperature,  allowing a  mass
estimation     through    a    $M_{500}-Y_{X}$     scaling    relation
\citep{vikhlinin09} with approximately 15\% accuracy.

From  X-ray $M_{500,Y_X}$ and  spectroscopic redshifts
  \citet{andersson10}  estimated   the  physical  $r_{500,Y_X}$,  both
  defined  with   respect  to  the  critical   density.   Here,  those
  $r_{500,Y_X}$/$M_{500,Y_X}$ are transformed to  $r_{200,Y_X}$/$M_{200,Y_X}$ using  the Navarro,
  Frenk, \&  White  \citep[][hereafter  'NFW' profile]{navarro97}
  radial  mass profile  with concentration  of 5  for the  dark matter
  halo,  which implies $r_{200,Y_X}=1.51\times  r_{500,Y_X}$ / $M_{200,Y_X}=1.38\times  M_{500,Y_X}$  conversion. The  angular  projection   is  calculated  using  the  spectroscopic
  redshifts.  $M_{200,Y_X}$  as well  as $r_{200,Y_X}$ are  listed in
  Table~\ref{tab:xclustermass}.

\begin{table*}
\begin{center}
\caption{X-ray    masses,   spectroscopic   redshifts    and   cluster
  parameters.} \small
\begin{tabular}{cccccccc}
\hline\hline\rule[-2mm]{0mm}{6mm}
   ID      &  $M_{200,Y_X}^a$     &$z$ &$r_{200,Y_X}^b$& $r_{200,Y_X}$ & $\xi^c$ & $c_{red\ gal}^d$ & $c_{all\ gal}^e$ \\
          &[$10^{14} M_{\odot}$]&         & [Mpc]&[arcmin]& S/N  &   & \\ 
\hline 
SPT-CL J0516-5430 & $16.38  \pm 1.72$  & 0.2952$^f$ & 2.21 & 8.34 & 9.42 & $4.65^{+0.81}_{-0.73}$ & $2.79^{+0.63}_{-0.52}$ \\  
SPT-CL J0509-5342 & $7.51  \pm 0.83$   & 0.4626$^g$ & 1.61 & 4.54 & 6.61 & $3.18^{+3.50}_{-1.39}$ & $1.94^{+7.44}_{-1.36}$\\ 
SPT-CL J0528-5300 & $4.11  \pm 1.19$   & 0.7648$^g$ & 1.17 & 2.61 & 5.45 & $5.93^{+5.78}_{-2.58}$ & $3.23^{+1.37}_{-0.55}$\\ 
SPT-CL J0546-5345 & $7.37  \pm 0.85$   & 1.0665$^h$ & 1.27 & 2.57 & 7.69 & $4.02^{+1.98}_{-1.37}$ & $4.04^{+1.92}_{-1.31}$\\
\hline
\end{tabular}
\tablecomments{$^a1.38\times    M_{500}$   from   \citet{andersson10},
  assuming   a  concentration   parameter  of   $c=5$  for   the  mass halo. 
  $^b1.51\times  r_{500,Y_X}$ from \citet{andersson10}, assuming a concentration parameter  of $c=5$ for the mass  halo.  
  $^c$The S/N measured in 150 GHz  SPT maps from from \citet{vanderlinde10}.  
  $^d$ Concentration parameter  from the NFW  fitting of the  red galaxies.
  $^e$  Concentration  parameter  from  the  NFW fitting  of  the  all galaxies.   
  $^f$Spectroscopic   redshift  from  \citet{bohringer04}.
  $^g$Spectroscopic  redshift  from \citet{high10}.  
  $^h$Spectroscopic redshift  from \citet{brodwin10}.}
\label{tab:xclustermass}
\normalsize
\end{center}
\end{table*}

\section{Cluster Galaxy Populations}

The  galaxy populations  in  clusters have  been  studied using  several
techniques  and selection  processes.  Clusters  of galaxies  have been
selected      mainly      from      optical     images      \citep[for
example]{abell58,abell89,koester07a} and  through their X-ray emission
\citep[among  others]{ebeling96,Vikhlinin98,bohringer04}.  A selection
of clusters based on  their Sunyaev-Zeldovich signature promises a less
biased  selection method,  as  it is  likely  to be  less affected  by
projections    or     false    clusters    like     optical    surveys
\citep{lucey83,sutherland88,collins95,cohn07}, and  its mass selection
function is nearly redshift independent, making the cluster sample more
homogeneous  than X-ray  surveys  in redshift  space.     Also, the  SPT
survey will be  able to find the most massive  clusters of galaxies in
the universe  \citep{carlstrom02,carlstrom09}.  Once  completed,  the size, redshift extent 
and the  degree of completeness of the SZE selected cluster sample 
will be ideal for statistical analysis of astrophysical properties in  high mass clusters.  Here  we focus  on the  first four  SPT selected clusters, which are all high mass systems extending over a broad redshift range.

\subsection{Radial distribution of galaxies}

The radial distribution of galaxies in clusters can be used to further 
our understanding  of the  cluster environment physics.   For example,
from  N-body and  gas dynamical  simulations, which  include radiative
cooling,   star   formation,    SN   feedback,   UV   heating,   etc.,
\citet{nagai05}   produced   radial   distributions  consistent   with
observations of X-ray selected cluster samples from \citet{carlberg97}
and \citet{lin04a}.  \citet{saro06}, using hydrodynamical simulations,
also  showed  an agreement  between  the  radial  distribution of  the
simulated  galaxies and  X--ray and  optically selected  clusters from
\citet{popesso07b}.  

For  the following analysis  we define  the cluster  center to  be the
position  of  the  observed  brightest  cluster  galaxy  member  (BCG;
coordinates are  listed in Table~\ref{tab:completeness}),  which agree
with  its  X-ray  center  \citep{andersson10}.  In  order  to  compare
different studies we estimate the concentration parameter from the NFW
surface  density  profile.   We  obtain  the NFW  surface  density  by
integrating    the    three-dimensional    number   density    profile
$n(x)=n_0x^{-1}(1+x)^{-2}$        along        the       line-of-sight
\citep[see][]{Bartelmann96}  where   $x=c_gr/r_{200}$,  $n_0$  is  the
normalization and  $c_g$ is  the concentration parameter.   On stacked
cluster data it is customary  to fit both parameters, $n_0$ and $c_g$.
In our case the  NFW fit is done over single cluster  data to a common
magnitude limit,  and this leads to considerable  uncertainties in the
parameters of the NFW profile.   In order to minimize this problem, we
introduce  the {\it  observed}  number of  galaxies  in the  equation.
Integrating the  NFW surface  density over the  projected area  we can
derive  $n_0=n_0(n_{obs},c_g)$  allowing us  to  fit  the NFW  density
profile  as a single  parameter function.   Also, after  a statistical
background correction,  a background  fitting is performed  along with
the NFW  fit. Such background fit is  limited within the
Poisson uncertainty of the observed background.

\begin{figure}
\begin{center}
\includegraphics[width=0.49\textwidth]{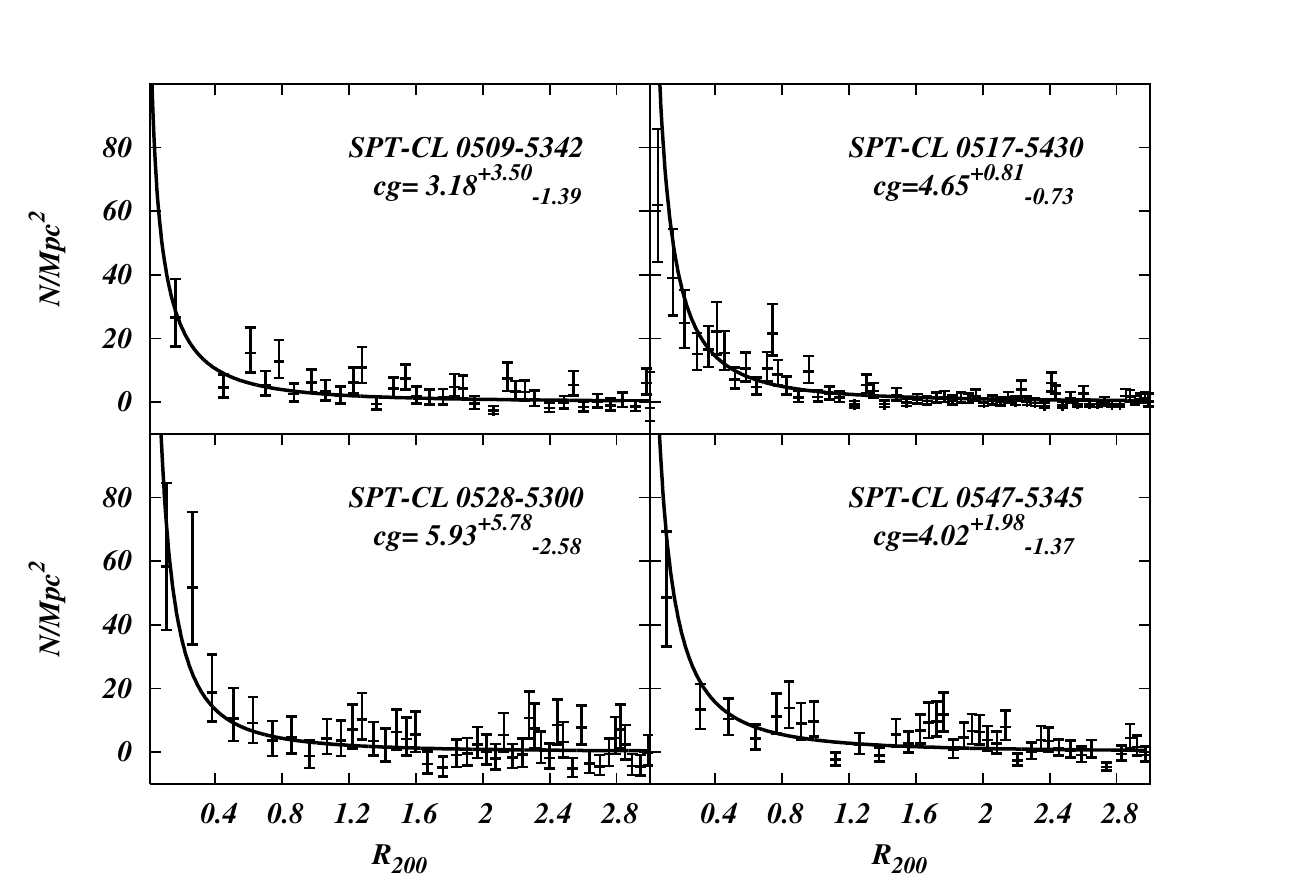}
\caption{Radial  profiles  for the  red
    galaxy population to $0.36L^*$  (same as blue fraction), binned to
    obtain similar  signal to noise.   These profiles are  centered on
    the BCG and  extend to 3$r_{200}$ to allow  the background and the
    cluster  profile to be  fit simultaneously.  All  radial profiles  are consistent
    with  NFW  profiles with  concentrations $c\sim4$.}\label{radial}
\end{center}
\end{figure}

 The  radial surface  density profiles  are constructed
  using both the red+blue and  the red galaxy population defined from the
  color-magnitude    diagram   of    the    red     sequence    (see
  sec.~\ref{sec:bluefraction}).    The  galaxy   population   is  also
  selected performing  a cut  in brightness, selecting  galaxies which
  are  fainter  than the  BCG  and brighter  than  a  common limit  of
  $0.36L^*$.  The error bars are computed using small number statistics
  \citep{gehrels86}.  The background is statistically subtracted and a
  second correction is  applied fitting it to a  radius of $3r_{200}$.
  Finally the  data is presented using radial  bins of constant signal-to-noise
  of 3.5 (see Fig.~\ref{radial}).

  Some corrections are applied to these profiles.   In the area
  calculation  for  each radial  bin  in  Fig.~\ref{radial}, the  area
  covered by saturated  stars was excluded in order  to avoid an under
  estimation of the surface  density.  This is especially important in
  the case of SPT-CL J0509-5342 where several bright stars close to the  BCG
  are blocking  the detection of  galaxy cluster members, covering about 50\% of the area at a radius of $0.4r_{200}$.

The     concentration     parameters     found    are     shown     in
Table~\ref{tab:xclustermass}.   With a concentration  in the  range of
$c\approx 3-6$,  the clusters agree at $1\sigma$  confidence.  We note
that the blue+red distribution tends  to be less concentrated than the
red population alone, which is consistent with previous analyses where
a   higher    concentration   is   seen   in    the   red   population
\citep[e.g.][]{goto04}.   The concentration  we find  is  in agreement
with concentrations  drawn from  X-ray selected clusters  of galaxies.
For     example,    \citet{carlberg97}     found     a    $c_g$     of
$3.70^{+3.99}_{-1.38}$ at 95\% confidence,  using 16 clusters from the
CNOC survey  with a median redshfit  of $\sim 0.3$ for  a similar mass
range     ($2\times     M_{\odot}^{14}$-$6.6\times    M_{\odot}^{15}$;
\citet{carlberg96}).  \citet{lin04a},  from stacked 2MASS  K-band data
on  93  nearby  X-ray  selected  clusters,  found  a  value  of  $c_g=
2.90^{+0.21}_{-0.22}$  in  a  wider $3\times  M_{\odot}^{13}$-$2\times
M_{\odot}^{15}$ range. Both are consistent with our results.

We also  found agreement  between our concentration  parameter and
the  concentration  parameter found  for  optical selected  clusters.
\citet{biviano09}  found,  studying  19  intermediate  redshift  ($0.4
\lesssim z  \lesssim 0.8$; $0.7 \lesssim M_{200}  \lesssim 13.6 \times
10^{14} M_\odot$) EDisCS+MORPHS clusters, a concentration parameter of
$c=3.2^{+4.6}_{-2.0}$. Also, \citet{johnston07}  found, using the SDSS
sample, a concentration parameter of $c_{200|14}=4.1\pm 0.2_{stat} \pm
1.2_{sys}$ for a cluster mass of $M=10^{14}h^{-1}M_\odot$.
In summary,  we find  no evidence that  SZE selected clusters exhibit different galaxy radial distributions than in optical and X-ray selected clusters.

\begin{figure}
\begin{center}
\includegraphics[width=0.49\textwidth]{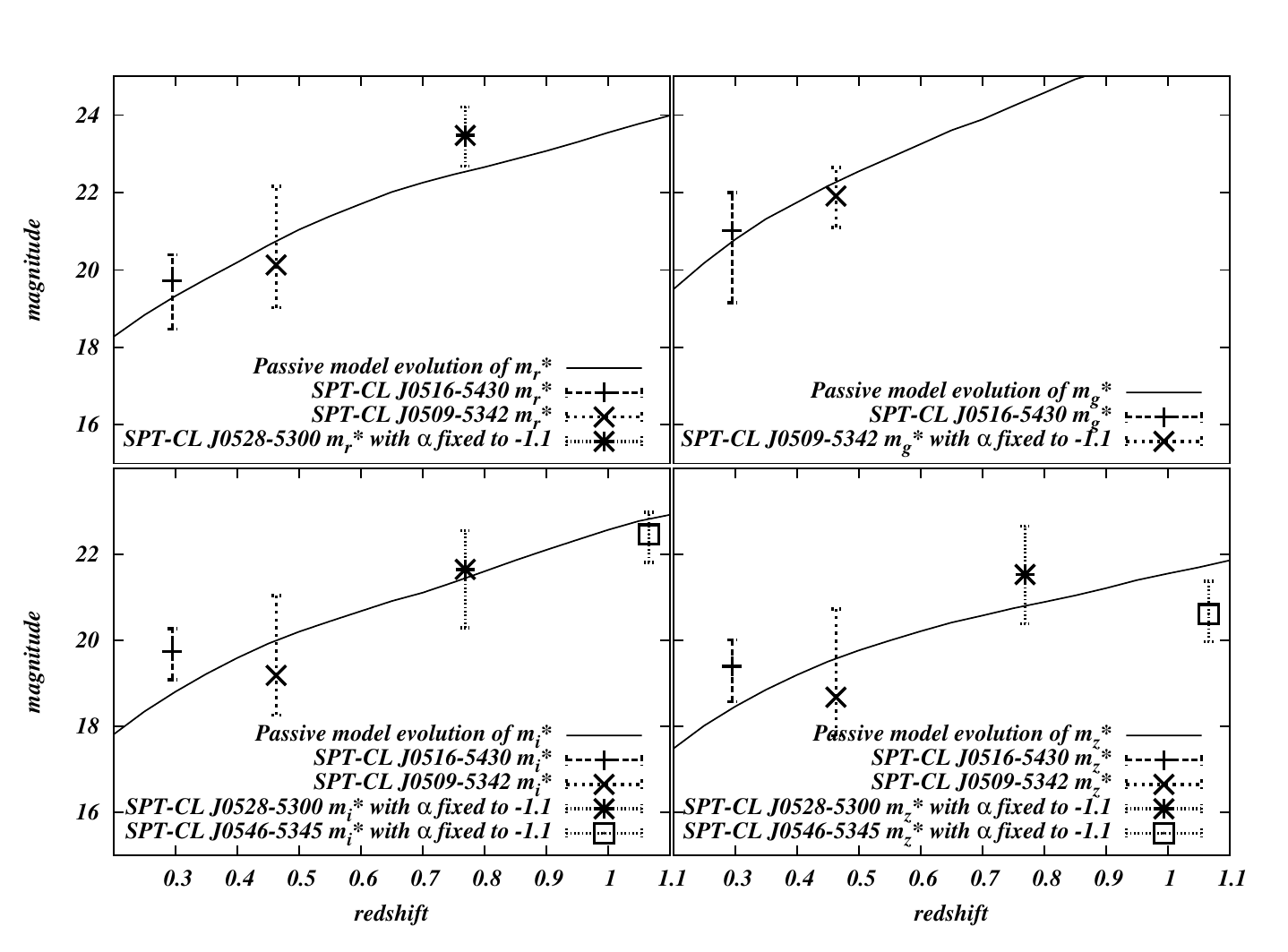}
\caption{Here we  plot the $m^*$  (with 1 $\sigma$  uncertainties) for
  each  band  that  results   from  Schechter  function  fits  to  the
  luminosity  function  with   free  parameters  $m^*$,  $\phi^*$  and
  $\alpha$ ($\alpha$ fixed where noted).   We limit  the range  of $m^*$  to be  fainter  than the
  identified BCG  for each cluster.  The continuous  line represents a
  passive    evolution   single    burst   model    at    $z=3$   from
  \protect\citet{bruzual03}.   It is  clear that  these SPT  selected clusters
  have  galaxy  populations  consistent  with  this  simple  evolution
  model. \label{evolution}}
\end{center}
\end{figure}
\newpage
\subsection{Luminosity functions}
\label{sec:lf}

The luminosity function (LF) is an important tool for testing theories
of  galaxy formation and  evolution.  For  example, ever  more complex
simulations  can  be  tested  against  the  LF,  as  an  observational
constraint, to probe our understanding of the evolution of galaxies in
the  cluster's environment  \citep{romeo05,saro06}.  With  clusters of
similar masses we can study the  LF as a function of redshift and with
the LF  parameters we can  calculate the Halo Occupation  Number (HON)
and test the \textit{N-M} scaling relation \citep{lin04a,lin06}.

The  LF can  be described  by the  three parameter  Schechter function
\citep{schechter76},
\begin{displaymath}
\phi(m)=0.4
\ln(10)\ \Phi^*10^{0.4(m^*-m)(\alpha+1)}\exp(-10^{0.4(m^*-m)})
\end{displaymath}
where  $\Phi^*$  is the  normalization,  $m^*$  is the  characteristic
magnitude and $\alpha$  accounts for the faint end  power law behavior
of the function.  

The construction of the LF is  done assuming that the $observed$ LF in
the  cluster area  is  the superposition  of  the cluster  LF and  the
background/foreground non--cluster  LF.  To recover the  cluster LF we
subtract  the galaxy  source count,  rescaled  by the  area, from  the
observed LF.   Given the wide range  in redshift we present  the LF in
the four $griz$ bands.

The area of the cluster is defined by our estimation of $r_{200}$ (see
\S\ref{sec:xmass} and Table~\ref{tab:xclustermass}), and the area of
the  background is the  tile area  (36'$\times$36') minus  the cluster
area.  The bright end limit of  the LF is defined by the cluster's BCG
while the  faint end limit is  defined by its completeness  at 90\% or
50\%,  depending on  the  redshift,  in each  band.   Below the  100\%
completeness, the 0.5 mag  bins are corrected using  the error
function  fitted to  the BCS/CFHT  comparison galaxy  count histograms
described in \S\ref{sec:completeness} and shown in Fig.~\ref{comp0}.
Finally, the  number of galaxies, background corrected,  is divided by
cluster volume (in M$pc^3$)  and the uncertainty is assumed Poissonian
in the total number of galaxies (cluster plus background).

Below  we extract  $m^*$  and  $\alpha$ from  our  cluster sample  and
compare them to previous results drawn from X-ray and optical selected
clusters of galaxies.

\subsubsection{Evolution of $m^*$}
\label{sec:mstar}

Studies of  $m^*$ evolution  in clusters have  been done  in different
wavelengths  and  with  different  selection methods.   These  studies
indicate that the stellar populations  in many of the cluster galaxies
have  evolved passively  after  forming at  high redshift  \citep[see,
  e.g.,][and                                                 references
  therein]{gladders98,delucia04,holden04,muzzin08}.  There are several
indications  that  $m^*$ evolution  can  be  described  from a  single
stellar  populations  (SSP)  synthesis  model  as  optical  and  X-ray
selected clusters.   All four  of these SPT  selected clusters  in our
study  show red  sequences  (see Fig.~\ref{bf_pop}),  and their  color
evolution  is consistent  with colors  derived from  a  single stellar
population (SSP) synthesis model.

In  order  to  perform  a  direct comparison  of  the  brightness  and
evolution of the characteristic magnitude, we let all the LF variables
vary, where possible,  and compare $m^*$ in $griz$  bands derived from
the LF fitting to  that based on the SSP model.  The  SSP model we use
for  the red  galaxy  population  is constructed  using  a Bruzual  \&
Charlot synthesis model  \citep[BC03;][]{bruzual03} for the red galaxy
populations, assuming a single burst of star formation at z=3 followed
by passive  evolution to  z=0.  We use  six different models  with six
distinct  metallicities  to match  the  tilt  of  the color  magnitude
relation   at   low   redshift,    and   we   add   scatter   in   the
metallicity-luminosity relation to  reproduce the intrinsic scatter in
the color-magnitude relation.  These models are then calibrated, using
51 X-ray clusters  that have available SDSS magnitudes  drawn from the
DR7 database.   Details of the model used  can be found in  Song et al
(submitted).   As shown  in  Fig.~\ref{evolution}, the  SSP model  and
$m^*$ in each  band are in good agreement, showing  that the SSP model
is an appropriate description of both the colors and the magnitudes of
the more  evolved early type galaxies  in this sample  of SZE selected
clusters.  We will use this  agreement to carry out a more constrained
study of the luminosity function.

\subsubsection{Faint end slope}
\label{sec:lfalpha}

To  learn  about the  $\alpha$  behavior  we  take advantage  of  the
agreement shown in \S~\ref{sec:mstar}  between SSP model and the data.
We adopt $m^*$  from the model (see Table~\ref{tab:lfun})  and fit for
$\Phi^*$ and $\alpha$ for each cluster individually.  The study of the
faint end slope $\alpha$ provides  us with information about the faint
galaxy populations  in the  cluster with respect  to the  more evolved
bright end, which is dominated  by luminous early type galaxies.  This
relation gives us insight into competing processes in the hierarchical
structure  formation  scenario,   including  the  accretion  of  faint
galaxies by the  cluster, causing a steep $\alpha$,  and the evolution
of  galaxies  inside the  cluster  through  galaxy merging,  dynamical
friction, star formation quenching and other processes.

\begin{figure}
\begin{center}
\includegraphics[width=0.49\textwidth]{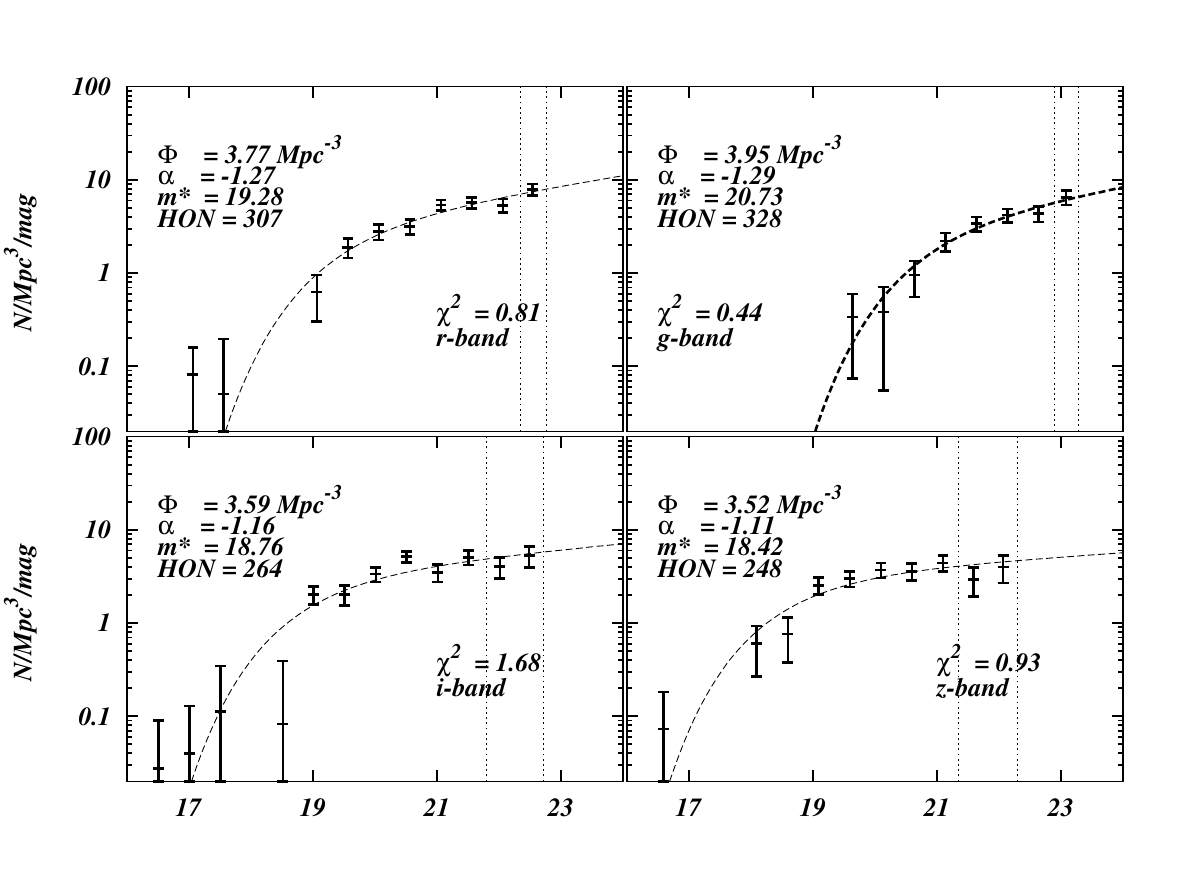}
\caption{Here we plot the luminosity function with best fit Schechter
  function for each  band in SPT-CL J0516-5430. Note  that the BCG had
  been removed.   Best fit parameters  are shown on the  figure, while
  Table~\ref{tab:lfun} includes best  fit and 1$\sigma$ uncertainties.
  100\%  and 90\% completeness  limits are  noted with  vertical dotted
  lines in each panel.\label{lf0}}
\end{center}
\end{figure}

\begin{figure}
\begin{center}
\includegraphics[width=0.49\textwidth]{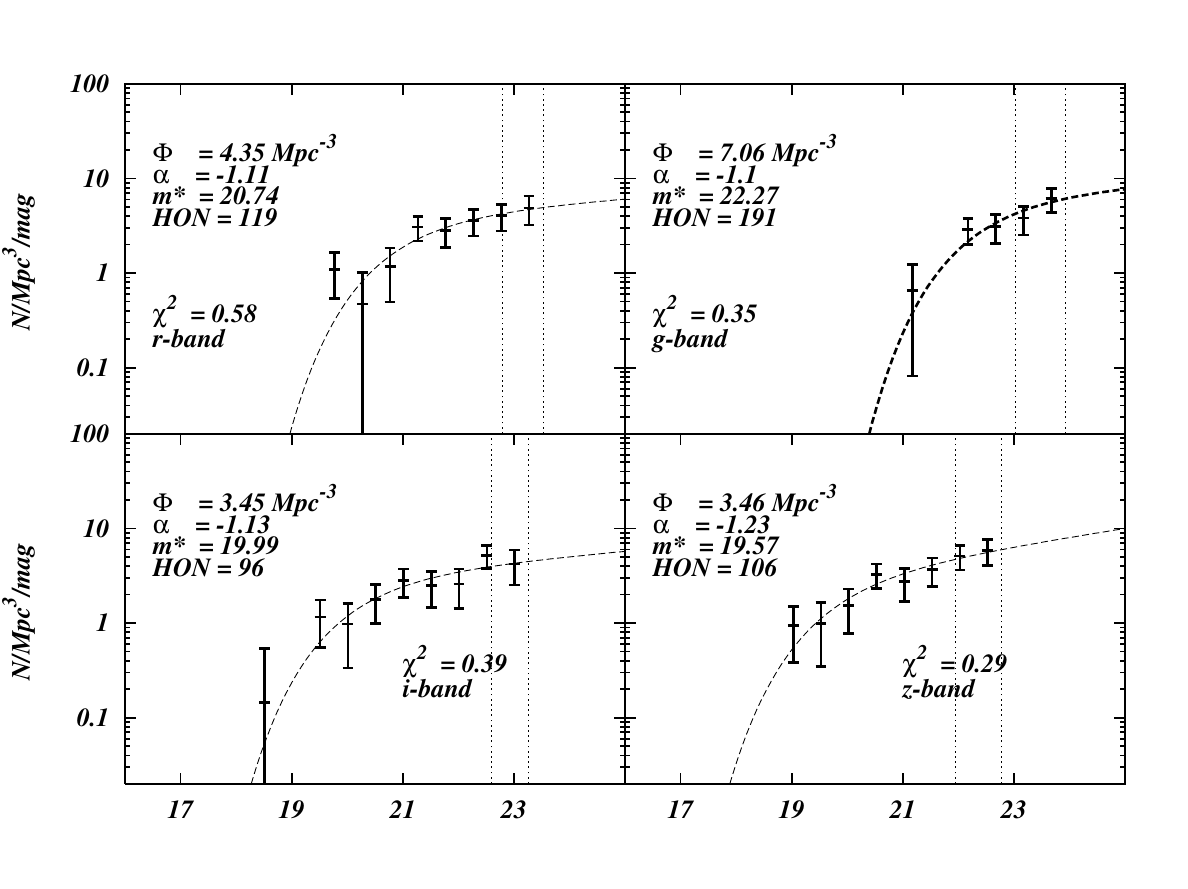}
\caption{Same     as    Figure~\ref{lf0}     but     for    SPT-CL
  J0509-5342.\label{lf9}}
\end{center}
\end{figure}

\begin{figure}
\begin{center}
\includegraphics[width=0.49\textwidth]{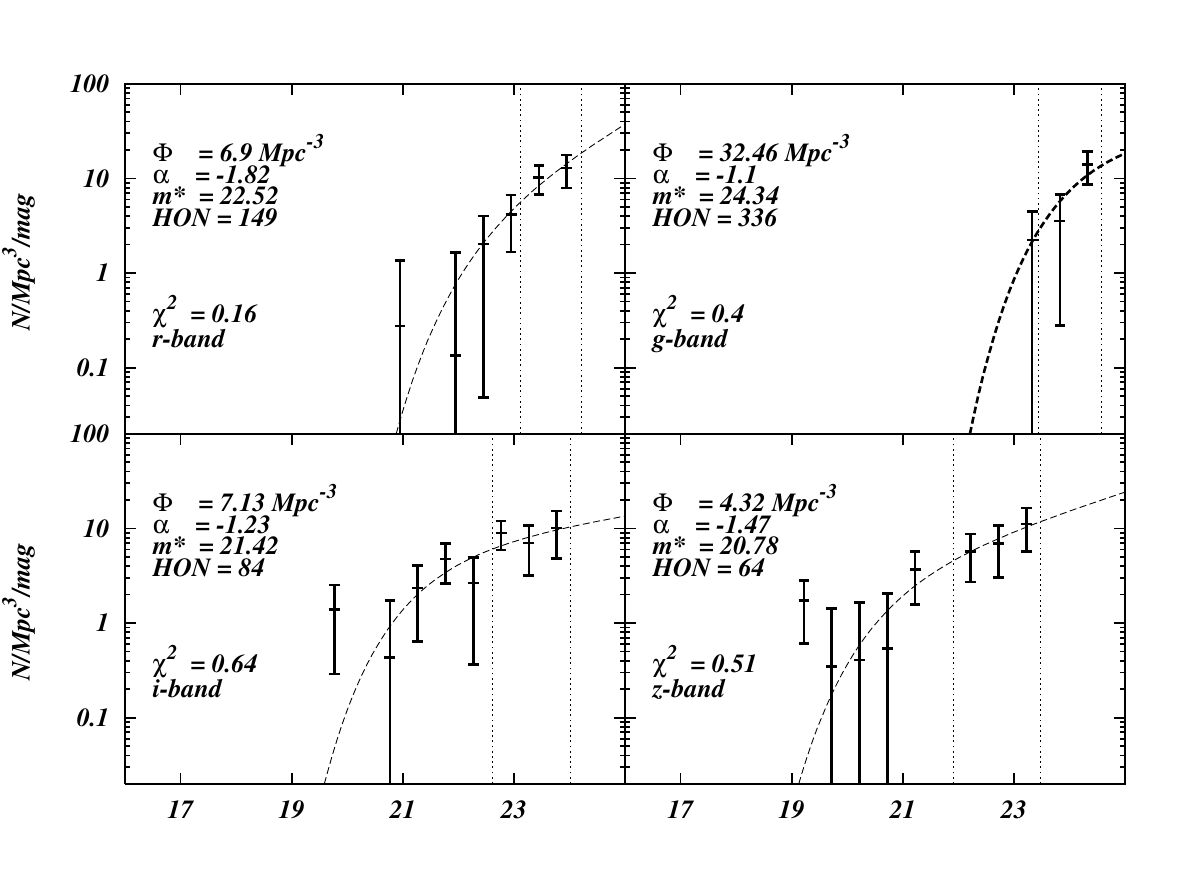}
\caption{Same as  Figure~\ref{lf0} but for SPT-CL  J0528-5300 with the
  100\% and 50\% completeness  limits noted with vertical dotted lines
  in each panel.\label{lf7}}
\end{center}
\end{figure}

\begin{figure}
\begin{center}
\includegraphics[width=0.49\textwidth]{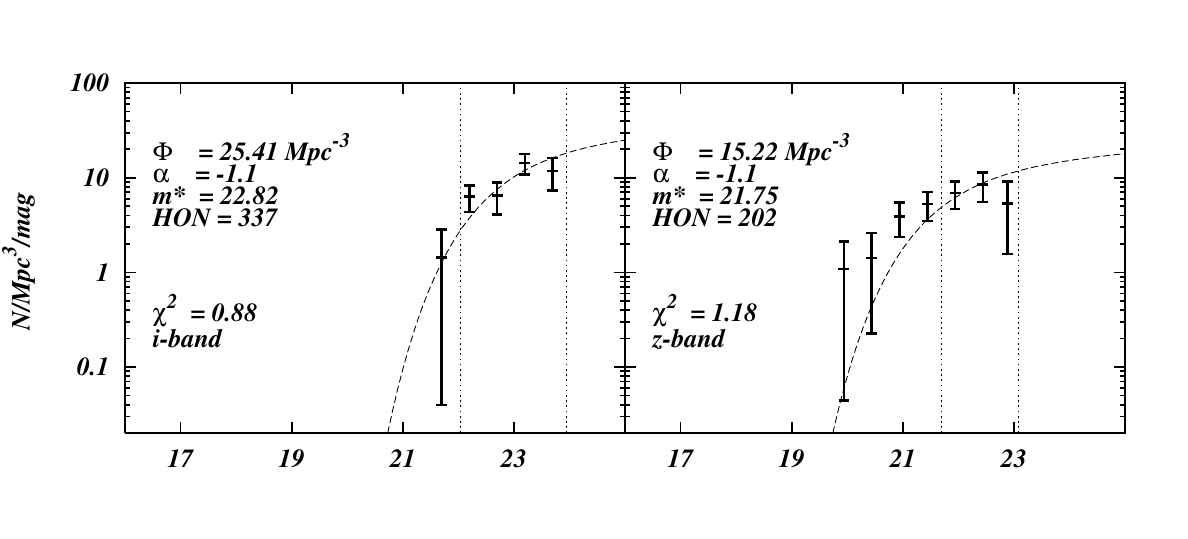}
\caption{Same     as    Figure~\ref{lf7}         for    SPT-CL
  J0546-5345 with only $i$ and $z$ bands. \label{lf1}}
\end{center}
\end{figure}

\begin{figure}
\begin{center}
\includegraphics[width=0.45 \textwidth]{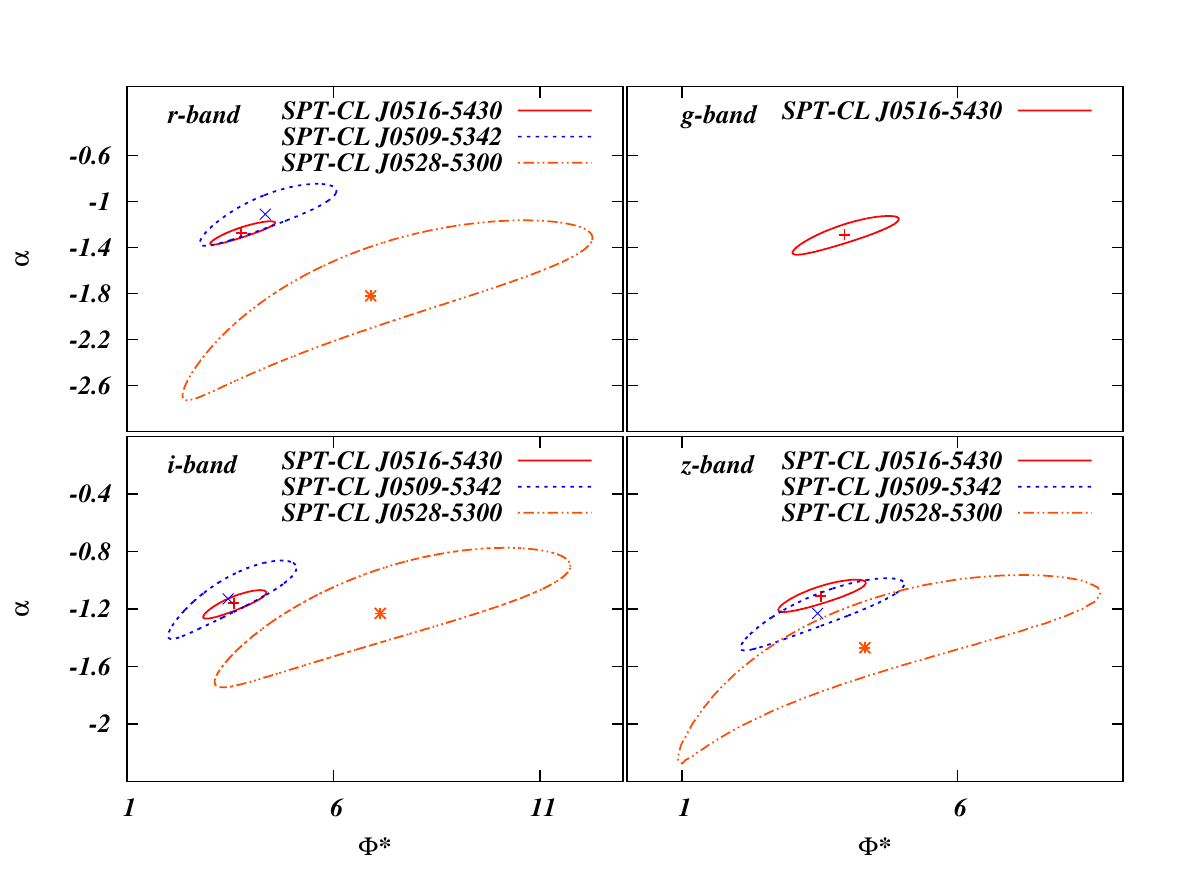}
\caption{We plot the 68\% confidence  region for the LF parameters for
  each cluster and band combination.  Panels are arranged by band with
  confidence regions  for each  cluster where a  fit for  $\alpha$ and
  $\Phi^*$ was possible.  The current data suggest steeper than normal
  faint end parameters $\alpha$ in two  of the clusters and there is a
  tendency   for  the   higher   redshift  systems   to  have   higher
  characteristic  galaxy   densities,  as  expected   in  an  evolving
  universe.   Values including those  at fixed  $\alpha$ are  given in
  Table~\ref{tab:lfun}.
\label{confidence}}
\end{center}
\end{figure}

Using  the  $amoeba$  simplex  minimization  routine  \citep{press92},
$\Phi^*$ and  $\alpha$ are chi-square fitted,  and their uncertainties
are  determined  by  gridding  in  parameter  space  (see  the  LF  in
Fig.~\ref{lf0},  \ref{lf9}, \ref{lf7},  \ref{lf1},  and their  contour
confidence regions at Fig.~\ref{confidence}).

From the literature, we find that our average $\alpha\approx$ -1.2 is
in  agreement at the 1$\sigma$  level with  previous  studies which  used
samples  constructed with different  selection methods.   For example,
from an  optical work, \citet{depropris03}  used 60 clusters at  z $<$
0.11  from  2dFGRS in the  $b_J$  band  finding  $\alpha=-1.28 \pm  0.03$.
\citet{paolillo01} found, on a
composite    LF   of    39    Abell   clusters,    an   $\alpha$    of
$-1.07^{+0.09}_{-0.07}$,          $-1.11^{+0.09}_{-0.07}$          and
$-1.09^{+0.12}_{-0.11}$ for Gunn $g$, $r$ and $i$ respectively.

From X-ray selected samples  \citet{lin04a} created a composite K-band
LF of 93 clusters, finding that  the faint-end slope is well fitted by
$-1.1 \lesssim  \alpha \lesssim -0.84$ in agreement  with our findings
within the  errors.  \citet{popessoII05}, using 97 X-ray selected
clusters with SDSS  photometry, for a redshift $z<$  0.25, found that a
better representation of the data is given by two Schechter functions,
characterized by  a bright  and a faint  end slope.  Comparing  to the
bright  end of  the double  Schechter function  with  local background
subtraction (which is the most  similar case), the bright end slope, in
1~Mpc $h^{-1}$, has a slope $\alpha$ of $-1.23\pm0.11$, $-1.05\pm0.13$,
$-1.17\pm0.13$,  and  $-1.06\pm0.12$  in   $g$,  $r$,  $i$,  and  $z$,
respectively, also agreeing with our findings at the 1$\sigma$ level.

From IR selected clusters \citet{muzzin08} detected 99
  clusters  and   groups of  galaxies  and  constructed  the  LF  in
  3.6$\mu$m,  4.5$\mu$m,   5.8$\mu$m,  and  8.0$\mu$m.   Although  the
  3.6$\mu$m band  is redward of our
  $griz$ photometry, the LF constructed seems to be consistent with $\alpha \approx -1$.

The agreement  found between  the multiband LF  parameters calculated
for our SZE selected clusters,  and previous studies of galaxy cluster
LFs  indicates  that the  galaxy  populations  in  these SZE  selected
clusters are  not very  different from those  in clusters  selected by
other means.

\begin{table*}
\begin{center}
\caption{HON parameters} \small
\begin{tabular}{ccccccccccc}
\hline\hline  \rule[-2mm]{0mm}{6mm}  
ID                & $\Phi_g^*$  & $\alpha_g$  & $m^*_g$ & $HON$      &$\chi^2$ &  $\Phi_r^*$ & $\alpha_r$ & $m^*_r$ & $HON$       &$\chi^2$ \\  
                  & $[Mpc^{-3}]$&             &         &$(m_g^*+2)$ &         & $[Mpc^{-3}]$ &           &         & $(m_r^*+2)$ &  \\ 
\hline 
SPT-CL J0516-5430 & $3.95^{+0.98}_{-0.94}$& $-1.29^{+0.16}_{-0.18}$& 20.73 & $327^{+157}_{-112}$& 0.44 & $3.78^{+0.81}_{-0.76}$ & $-1.27^{+0.10}_{-0.10}$ & 19.28 & $308^{+105}_{-83}$ & 0.81\\ 
SPT-CL J0509-5342 & $7.07^{+1.11}_{-1.11}$& -1.1$^a$             & 22.27 &  $191^{+30}_{-30}$ & 0.35 & $4.36^{+1.71}_{-1.59}$ & $-1.11^{+0.26}_{-0.28}$ & 20.74 & $119^{+95}_{-57}$  & 0.58\\ 
SPT-CL J0528-5300 & ---                 & ---                  & 24.34 & ---              &  --- & $6.89^{+5.37}_{-4.54}$  & $-1.82^{+0.66}_{-0.91}$ & 22.52 & $149^{+691}_{-122}$& 0.16\\ 
SPT-CL J0546-5345 & ---                 & ---                  & 25.92 & ---              &  --- & ---                   &  ---                  & 23.85 &  ---              & ---\\ 
\hline 
ID                & $\Phi_i^*$  & $\alpha_i$ & $m^*_i$ & $HON$      &$\chi^2$ & $\Phi_z^*$   & $\alpha_z$ & $m^*_z$& $HON$      &$\chi^2$ \\ 
                  & $[Mpc^{-3}]$&            &         &$(m_i^*+2)$ &         & $[Mpc^{-3}]$ &            &        & $(m_z^*+2)$&         \\ 
\hline 
SPT-CL J0516-5430 & $3.59^{+0.78}_{-0.74}$ & $-1.17^{+0.10}_{-0.10}$ & 18.76 & $266^{+88}_{-71}$ & 1.67 & $3.52^{+0.82}_{-0.77}$ & $-1.11^{+0.11}_{-0.11}$ & 18.42 & $248^{+89}_{-71}$ & 0.93 \\
SPT-CL J0509-5342 & $3.45^{+1.65}_{-1.40}$ & $-1.13^{+0.27}_{-0.28}$ & 19.99 &  $96^{+87}_{-49}$ & 0.39 & $3.45^{+1.58}_{-1.38}$ & $-1.23^{+0.24}_{-0.26}$ & 19.57 & $105^{+92}_{-53}$ &  0.29\\
SPT-CL J0528-5300 & $7.15^{+4.57}_{-4.02}$ & $-1.23^{+0.46}_{-0.52}$ & 21.42 & $84^{+148}_{-58}$ & 0.64 & $4.34^{+4.25}_{-3.46}$ & $-1.47^{+0.50}_{-0.82}$ & 20.78 & $65^{+265}_{-56}$ & 0.51 \\
SPT-CL J0546-5345 & $25.43^{+4.03}_{-4.04}$& -1.1$^a$              & 22.83 & $337^{+53}_{-53}$ & 0.88 & $15.2^{+5.57}_{-5.58}$ & -1.1$^a$              & 21.75 & $202^{+74}_{-74}$ & 1.19\\
\hline
\end{tabular}
\tablecomments{$^a\alpha$  set to  fixed  value. $m^*$  from model  of  passive evolution.}\label{tab:lfun} 
\normalsize
\end{center}
\end{table*}

\subsection{Halo Occupation Number}
Based on  the Press \&  Schechter formalism \citep{press74},  the halo
occupation  distribution  (HOD)  is  a powerful  analytical  tool  for
understanding   the  physical   processes  driving   galaxy  formation
\citep{seljak00a,berlind03}.  Also  the HOD  can be used  to constrain
cosmological parameters \citep{zheng07}.

One  of  the  key  ingredients   in  the  HOD  formalism  is  $\langle
N\rangle(M)$,  the  mean number  of  galaxies  per  halo or  the  Halo
Occupation  Number  (HON).   In the hierarchical scenario,  the HON  is  expected  to
increase slower than the mass. While a fraction of the accreted galaxies merged, the 
galaxy production becomes less efficient as larger haloes are also hotter and less efficient 
in gas cooling \citep{cole00}.
Observationally,  several   studies  with  cluster   samples  selected
optically  and  through  their  X-ray emission  have  been  performed,
reinforcing  that picture.   For  example, from  samples of  optically
selected  clusters  and  groups,  \citet{marinoni02}  found  $N\propto
M^{0.83\pm0.15}$ for systems with $M\gtrsim10^{13}h_{75}^{-1}M_\odot$.
Also, \citet{muzzin07} found $N_{500}\propto M_{500}^{0.71\pm0.11}$ in
the $\sim2\times10^{14}M_\odot- 2\times 10^{15}M_\odot$ mass range. In
the X-ray  selection method counterpart, \citet{lin04a}  found, from a
sample of  93 nearby clusters and  groups, $N\propto M^{0.87\pm0.04}$.
Combining  X-ray and  optically selected  clusters, \citet{popesso07b}
found $N\propto  M_{200}^{0.92\pm0.03}$.  A similar  picture was found
by \citet{rines04},  who used nearby X-ray luminous  Abell clusters of
mass   $\sim    3\times10^{14}h^{-1}M_\odot$   and   found   $N\propto
M^{0.74\pm0.15}$.

Here  we test  whether the  HON of  SZE selected  clusters  exhibits a
$N\propto  M^{\beta}$,   with  $\beta<1$,  behavior   shown  by  other
selection methods.

Due to the  small sample presented here, our  approach is to construct
the HON and  compare our results to the  \textit{N-M} scaling relation
and  evolution  constraints  obtained by  \citet{lin04a,lin06}.   That
scaling relation is appropriate in this analysis as it covers the mass
and  redshift range  of this  SZE  sample.  The  scaling relation  was
constructed using  X-ray selected  clusters in the  $3\times10^{13}M_\odot$ -
$2\times10^{15}M_\odot$ mass  range using  nearby clusters with  2MASS K-band
photometry, and  later, \citet{lin06}, counting galaxies to the depth m*+2, expanded the study  to the 0-0.9
redshift range showing that the relation does not strongly evolve.

The \citet{lin04a} \textit{N-M} relation is,
\begin{displaymath}
N_{200} =  (36 \pm 3)(M_{200}/(10^{14}h^{-1}_{70}  M_\odot))^{0.87 \pm
  0.04}
\end{displaymath}
To calculate $N_{200}$ we integrate the cluster luminosity function to
$L(m_{MODEL}^*+2)$  using the parameters  of the  Schechter luminosity
function   fit,  $\phi_*$,   L$_*$   and  $\alpha$   computed  in   \S\ref{sec:lf}. The total number of galaxies is
\begin{displaymath}
N= 1 +  N^s, \ with\ N^s =  V \phi_*\int^{\infty}_{y_{low}} y^{\alpha}
e^{-y}\ dy
\end{displaymath}
where the 1 comes from the BCG, which is not part of the LF fitting, V
is the  cluster volume,  and $y_{low}=L_{low}/L_*$.  We use the derived $M_{200,Y_X}$ masses
and uncertainties as explained in \S\ref{sec:xmass}   from \citet{andersson10}  Chandra   and  XMM  observations.   The
uncertainty in  $N_{200}$ is estimated  by propagating the  1 $\sigma$
uncertainty in $\phi_*$ and $\alpha$ through the integration of the LF
to $m^*+2$.

\begin{figure}
\begin{center}
\includegraphics[width=0.49\textwidth]{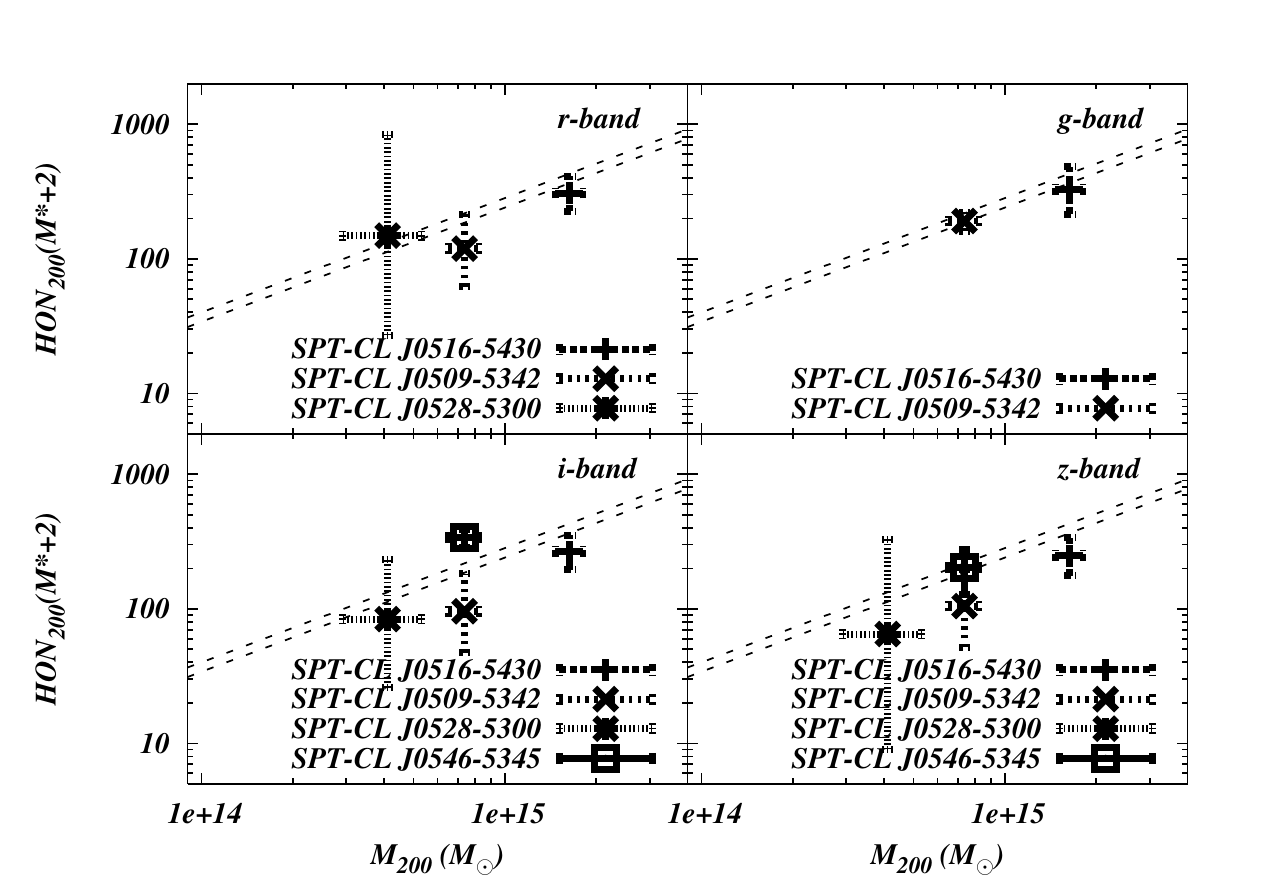}
\caption{We present  the halo occupation  number (HON(m$^*$+2)) within
  each band  for each cluster  where the LF  is measured in  more than
  three  bins. Masses and  uncertainties on  the horizontal  axis come
  from  X-ray analysis  of  Chandra observations  \protect\citep{andersson10}.
  HON uncertainties are  derived from the variation of  HON due to the
  1$\sigma$ uncertainty in the  LF ($\alpha$ and $\Phi$*).  The dotted
  lines show the HON derived from  a K-band analysis of a large sample
  of X-ray selected clusters \protect\citep{lin04a,lin06}.  These SPT selected
  clusters appear to be neither richer nor poorer.\label{HON}}
\end{center}
\end{figure}

The $N_{200}$ with their X-ray mass  for the four clusters in the four
observed band,  along with the  HON relation found  by \citet{lin04a},
are shown in Fig.~\ref{HON}.  Agreement between these SPT clusters and
the published results  on the X-ray selected sample  is good.  As with
the  concentration and  the  LF  faint end,  there  is no  significant
evidence  that  the  galaxy   properties  differ  from  those  already
extracted from previous X-ray selected cluster samples.

\subsection{Blue fractions}
\label{sec:bluefraction}

Another  property  of  the  galaxy  populations used  to  study  their evolution in clusters of galaxies is the blue fraction ($f_b$).
In  their  seminal  work  \citet[][BO hereafter]{butcher84},  using  a
samples of 33 optically selected clusters of galaxies, estimated $f_b$
and  showed  that  it   increased  with  look-back  time  (termed  the
Butcher-Oemler  effect).  Later  studies, such  as  \citet{rakos95} (0
$<z<$  1) and  \citet{margoniner00}  (0.03$\lesssim z  \lesssim$0.38),
using optically  selected clusters, also have found  a strong increase
in $f_b$ with redshift.

With the advent of new  optical surveys with hundreds or thousands of clusters
the analyses have been  strengthened statistically.  Using a sample of
$\approx$1000  clusters,  in a wide   redshift range (0$\lesssim  z \lesssim$0.9), 
drawn from the  Red-Sequence Cluster Survey (RCS), \citet{loh08}
found  a  mild correlation  between  the  red  fraction and  redshift.
\citet{hansen09}, using  thousands of  clusters and groups  from SDSS,
found  an  evolving $f_b$  in  the  two  redshift bins  studied (0.1-0.25 and 0.25-0.3),  also
noticing  that $f_b$  evolution was  weaker for  optical  masses above
$10^{14}h^{-1}M_\odot$.

Studies  using   samples  of   X-ray  selected  clusters,   have  been
contradictory.   \citet{kodama01}  used   a  sample  of  seven  clusters,  in the  redshift range of  0.23-0.43, and  found a
blue fraction trend consistent with BO, while \citet{fairley02}, using
a sample  of eight  clusters  in a  0.23-0.58 redshift
range found  virtually no trend with redshift.   
More  recently,  \citet{urquhart10}  used  CFHT MegaCam  $g$  and  $r$
photometry on 34 X--ray selected clusters in the redshift range  0.15-0.41 
to  study  $f_b$ correlation  with  other intrinsic  cluster
properties,  found  that  $f_b$  correlated with  mass  ($T_X$)  and
redshift.

Also   there    are   environmental   factors    to   be   considered.
\citet{smail98} used  10 X-ray  selected clusters at  similar redshift
(0.22 to  0.28) and found a  low blue fraction  of $\overline{f_b}= 0.04
\pm 0.02$ with a variation of $\Delta f_b = 0.06$, explained by 'small
accretion  events'  which  contribute  blue members  to  the  clusters
without  much increase  of  other  parameters such  as  mass or  X-ray
luminosity.  Such events could be  a source of scatter in the galaxy 
populations of clusters selected by any selection
method.   To analyze  $f_b$ correlation  with  other cluster
parameters  \citet{depropris04} used  a  sample of  clusters from  2dF
Galaxy Redshift Survey (2dFGRS) at  redshift $<$ 0.11, finding a large
variation ($f_b \sim$ 0.1-0.5  for $M^*+1.5$ at $r_{200}$) from cluster
to cluster.

The  apparent  contradiction  between  X-ray  and  optically  selected
samples and the sensitivity to environmental effects, raises questions
about how much of the observed $f_b$ is due to a selection method, how
much it is due to the  intrinsic scatter, and if these two effects can
conspire to produce an apparent trend where no trend exists.

What is needed  is a sample of galaxy clusters  which possess two main
characteristics: (1) the  selection of clusters is made  in a way that
is independent  of the  quantity whose evolution  is being  studied to
avoid possible  bias \citep{newberry88,andreon99}, and  (2) the sample
must contain  the same  class of clusters  (i.e.  same mass  range) at
different  redshift to help  in separating  mass trends  from redshift
evolution  \citep{andreon99}.  A  sample of  SZE selected  clusters of
galaxies  fulfills  these  requirements.   The selection  of  the  SZE
clusters  is closely  related  to  mass, and  that  mass selection  is
approximately independent  of redshift,  allowing a comparison  of the
same type of clusters at different epochs.

\begin{figure}
\begin{center}
\includegraphics[width=0.49\textwidth]{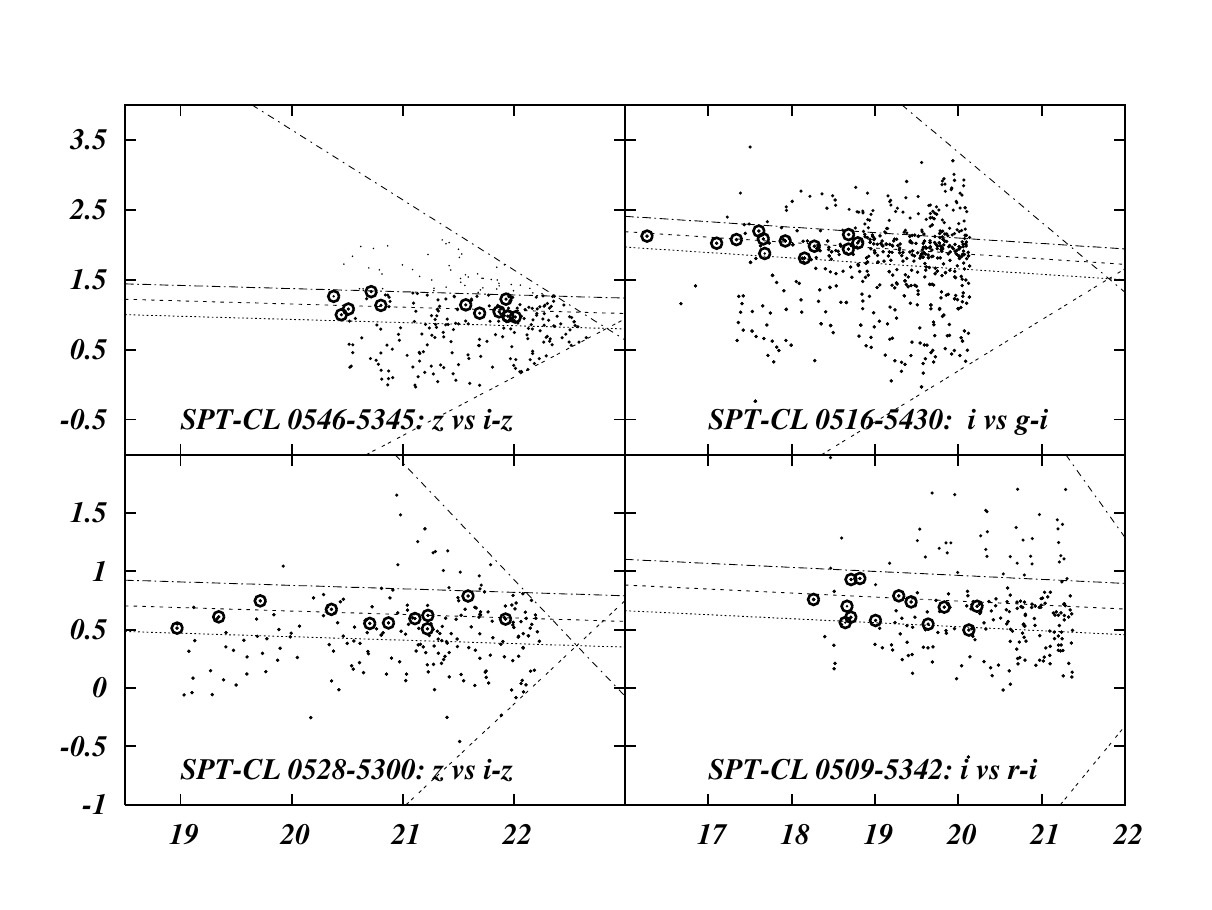}
\caption{Color magnitude diagram for galaxies around each cluster.  The blue population is defined to be more than 0.22~mags bluer than the red sequence.  Selection in magnitude uses the BCG  on the bright end and $m^*+\delta$ on the faint end, where this limit corresponds to the 90\%  completeness limit for SPT-CL~J0528-5300 ($0.36L^*_{MODEL}$).  The visually identified red sequence cluster galaxies are shown using circles.\label{bf_pop}}
\end{center}
\end{figure}

Historically $f_b$ has been measured in different ways. Initially the
average color of  the E/S0 galaxies, within a  radius of R$_{30}$ from
the  cluster center  that  is the  radius  that contains  30\% of  all
galaxies that  belong to the  cluster, and a concentration  index, were
use                  to                  define                  $f_b$
\citep[see][]{butcher84,rakos95,margoniner00,fairley02}.        Another
approach is using the red sequence from the color-magnitude diagram of
the   clusters  and   $r_{200}$  \citep{popesso07,barkhouse07}   or  a
combination of both methods, that is using the color-magnitude diagram
but R$_{30}$ \citep{kodama01,fairley02}.

Here we follow the approach of  using the red sequence to define the
red and  blue populations, and  $r_{200}$ to define the  radial extent.
This  ensures we  are using  the same  portion of  the  cluster virial
region, independent of redshift, and that we are exploring populations
with colors defined with respect to a passively evolving SSP model.

The galaxies used for the  $f_b$ measurement are inside the $r_{200}$
cluster  radius  and  are  fainter  than the  BCG  and  brighter  than
$0.36L^*_{MODEL}$.  They  are classified as red if  they are located
within $\pm$3 times the average dispersion of the Gaussian fit to the
color-magnitude relation \citep[$\pm$ 0.22 mag;][]{lopez04}, and blue
if they  are more than 0.22~mags bluer  than the Red  Sequence.  We  choose a
limit  of $0.36L^*_{MODEL}$  to allow  a meaningful  comparison among
three of our four clusters, as it is the deepest magnitude that we can
detect with good completeness for the three of them.  For the fourth cluster,
SPT-CL J0546-5345, we currently do not have  deep enough photometry for this analysis.

The color magnitude  diagram used for the clusters  depends on the red
sequence identification: g-i/i for SPT-CL J0516-5430, r-i/i for SPT-CL
J0509-5342, and i-z/z for SPT-CL J0528-5300 and SPT-CL J0546-5345 (see
fig. ~\ref{bf_pop}).  The blue fraction is defined as the statistically
background  corrected number  of blue  galaxies $n_b$  divided  by the
total  number of  statistically background  corrected  galaxies $n_t$.
The blue fraction and its gaussian propagated uncertainty are:
\begin{equation}
f_b=\frac{n_b}{n_r+n_b};\ \ \sigma_{f_b}^2=\sum_{i=r,b}\left(\frac{\partial{f}}{\partial{n_i}}\right)^2\sigma_{n_i}^2
\end{equation}
Where $n_b$ and $n_r$ are the blue and red statistically background
subtracted number of galaxies:
\begin{displaymath}
n_{i}=N_{i}-\overline{N}_{i}^{(bkg)}
\end{displaymath}
The uncertainties are expressed as
\begin{displaymath}
\sigma_{n_i}^2= \sigma_{N_i}^{2}+\sigma_{\overline{N}_i^{(bkg)}}^2
\end{displaymath}
assuming  $\sigma_{N_i}$  Poissonian.   The  last term  is  calculated
directly by measuring the RMS of the Gaussian distribution observed on
histograms  constructed from  the blue  and red  (or total)  number of
galaxies background corrected  in a circle of radius  $r_{200}$ on $n$
random       position      outside       the       cluster      radius
(background($r_{200}$)-$\overline{background}$)  in  order to  account
for background variations on  the observed 36'$\times$36' patch of the
sky.

A special mention  for SPT-CL 0509-5342 is required.  In the center of
the  cluster  are  three  bright  stars leaving  only  a  few  visible
galaxies; we  have corrected  this effect by  accounting for  the area
masked around  these stars.  Nevertheless,  the statistical background
subtraction   leads   to   negative   blue  galaxy   counts   in   the
$\sim0.6r_{200}$ inner part of the cluster area. 

\begin{figure}
\begin{center}
\includegraphics[width=0.49\textwidth]{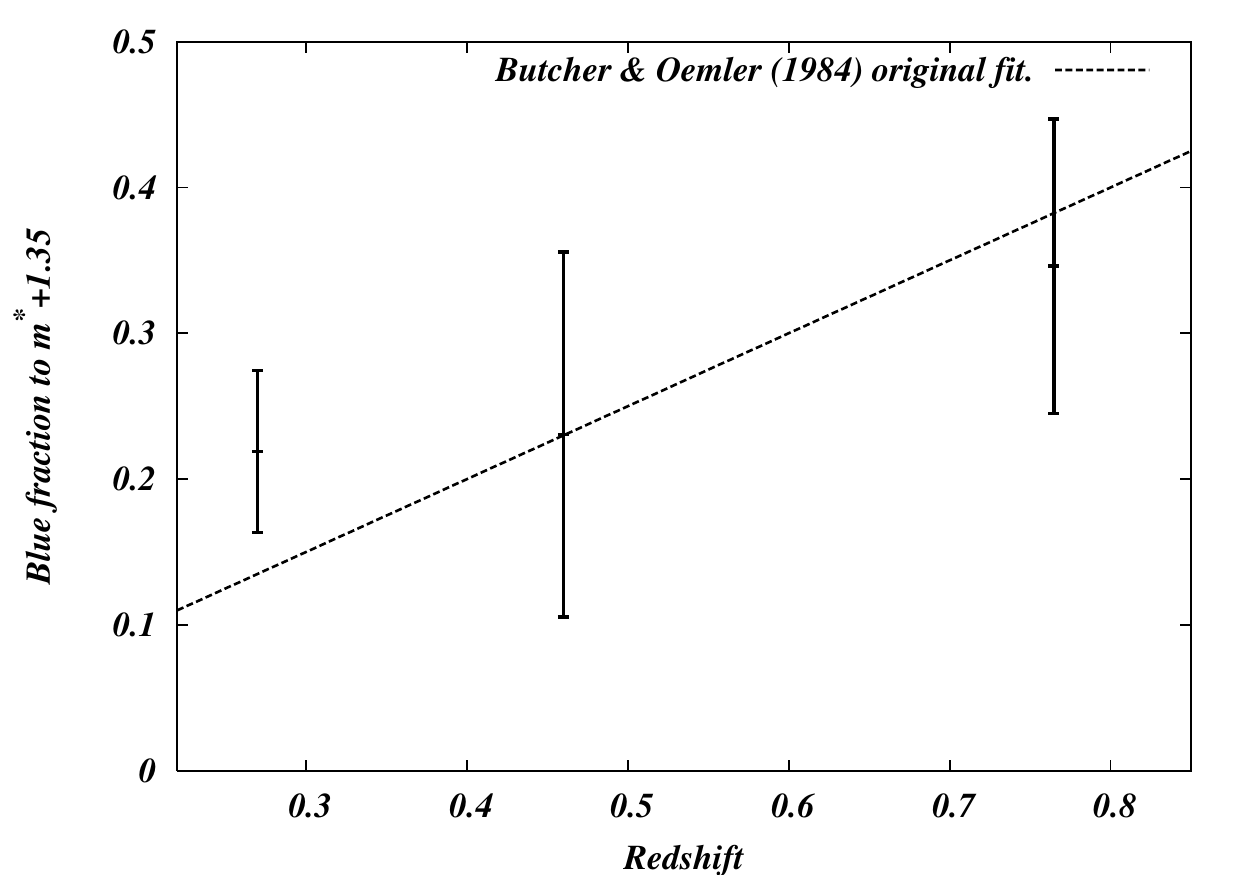}
\caption{Blue fraction versus redshift  using  the   populations  shown   in  Figure~\ref{bf_pop}.  The \protect\citet{butcher84} relation is shown (dashed line).  The  blue fraction is calculated using statistical background correction within $r_{200}$  and to a depth of $0.36L^*_{MODEL}$  for each system.\label{fb}}
\end{center}
\end{figure}

The blue fraction  of three of the four  clusters, at redshifts 0.295,
0.463 and 0.763, are shown in Fig.~\ref{fb}.  The measurements suggest
an increase  with redshift, as shown for  optically selected clusters,
although the result could be  consistent with a constant blue fraction
over the range  of redshift that we explored  with our limited sample.
Future  optical follow up  of SPT-SZE  selected clusters  using larger
aperture  telescopes on  the high  redshift end  will be  necessary to
understand the Butcher-Oemler effect in this cluster mass range within
the SZE selected sample.

\section{Conclusions}
\label{sec:conclusions}

We  present the  results of  a  careful examination  of the  multiband
optical properties  of the  galaxy populations in  the first  four SZE
selected galaxy clusters.  This  analysis builds upon the selection by
the South Pole Telescope survey,  deep multiband optical data from the
Blanco Cosmology Survey, Chandra  and XMM mass estimates and published
spectroscopic redshifts.

The  radial  distributions  of   galaxies  in  the  four  systems  are
consistent  with NFW profiles  with low  concentration in  the 2.2-3.6
range, although  the constraints in our highest  redshift clusters are
weak due  to the  imaging depth.  One  system shows a  clear secondary
peak, which  is evidence of multiple galaxy  components.  The observed
galaxy concentrations  in these SPT systems are  consistent with X-ray
and optical selected cluster samples as well as simulations.

We  showed that the  characteristic luminosities  in bands  $griz$ are
consistent with passively evolving  populations emerging from a single
burst at redshift $z=3$.  This is observed by direct comparison of the
$griz$  $m^*$ measurements  with  the evolution  of  the red  sequence
expected from the SSP model.

The  slope of  the luminosity  function, $\alpha$,  in all  four bands
showed an average of $-1.2$ consistent with previous studies and roughly
independent  of redshift, although  in the  high redshift  systems the
constraints are  weaker and the  $\Phi-\alpha$ contours are  much more
extended (see Fig.~\ref{confidence}) due to the depth of the data.

Halo occupation  numbers (to $m^*+2$)  for these systems appear  to be
consistent  with  the  relation  measured  in  X-ray  selected
clusters.  As  shown previously  \citep{lin04a}, this well  behaved and
simple galaxy  populations is unfortunately not  easy to use  as a mass
indicator  with  optical data  alone,  because  the  HON varies with the 
adopted virial radius of the cluster.

The blue fractions $f_b$ observed in these systems are consistent with
those  seen in  clusters  selected using  other  means.  Although  the
meassured  $f_b$  suggest a  redshift  evolution  (as optical  studies
show), it is within the  errors also consistent with a constant $f_b$.
It is clear that definitive  conclusions should be drawn with a larger
number of  clusters for more  robust statistics.  A larger  sample and
deeper multiband data on the high redshift end is needed.

The SPT selection provides  a powerful means
of  choosing similar  mass systems  over  a broad  range of  redshift,
making the  future larger cluster sample  particularly interesting for
this study.

In summary, our  systematic analysis of the galaxy  populations in the
first  SZE  selected  galaxy  clusters  spanning  the  redshift  range
$0.3<z<1.1$ provides no clear  evidence that the galaxy populations in
these SPT  selected clusters differ from populations  studied in other
X-ray and optically selected samples.  An extension of our analysis to
the full SPT sample will enable  a more precise test of the effects of
selection.  In addition, comparison  of the observed properties of the
SPT  cluster  galaxy  populations  and their  evolution  to  numerical
simulations of  galaxy formation should  allow for clean tests  of the
range of  physical processes that  are responsible in  determining the
formation and evolution of cluster galaxies.

\acknowledgments The South Pole Telescope is supported by the National
Science Foundation through grant  ANT-0638937. Partial support is also
provided by the  NSF Physics Frontier Center grant  PHY-0114422 to the
Kavli Institute of Cosmological  Physics at the University of Chicago,
the Kavli Foundation, and the  Gordon and Betty Moore Foundation.  AZ,
JM,  GB and  JL  acknowledge  the support  of  the Excellence  Cluster
Universe  in Garching.   This paper  includes data  gathered  with the
Blanco 4-meter  telescope, located at the  Cerro Tololo Inter-American
Observatory  in Chile,  which is  part  of the  U.S. National  Optical
Astronomy  Observatory,  which  is  operated  by  the  Association  of
Universities for Research in Astronomy (AURA), under contract with the
National  Science   Foundation.   This  work  is  based   in  part  on
observations obtained with the Chandra X- ray Observatory (CXO), under
contract SV4-74018, A31 with the Smithsonian Astrophysical Observatory
which operates the CXO for NASA.  We are very grateful for the efforts
of the  Chandra, XMM, and CTIO  support staff without  whom this paper
would not be possible. Support for M.B. was provided by the W. M. Keck
Foundation. B.S. acknowledges support from the Brinson Foundation.

\bibstyle{apj} \bibliography{ClusterOptical09.bbl}

\begin{thebibliography}{96}
\expandafter\ifx\csname natexlab\endcsname\relax\def\natexlab#1{#1}\fi

\bibitem[{{Abell}(1958)}]{abell58}
{Abell}, G.~O. 1958, \apjs, 3, 211

\bibitem[{{Abell} {et~al.}(1989){Abell}, {Corwin}, \& {Olowin}}]{abell89}
{Abell}, G.~O., {Corwin}, Jr., H.~G., \& {Olowin}, R.~P. 1989, \apjs, 70, 1

\bibitem[{{Andersson} {et~al.}(2010){Andersson}, {Benson}, {Ade}, {Aird},
  {Armstrong}, {Bautz}, {Bleem}, {Brodwin}, {Carlstrom}, {Chang}, {Crawford},
  {Crites}, {de Haan}, {Desai}, {Dobbs}, {Dudley}, {Foley}, {Forman},
  {Garmire}, {George}, {Gladders}, {Halverson}, {High}, {Holder}, {Holzapfel},
  {Hrubes}, {Jones}, {Joy}, {Keisler}, {Knox}, {Lee}, {Leitch}, {Lueker},
  {Marrone}, {McMahon}, {Mehl}, {Meyer}, {Mohr}, {Montroy}, {Murray}, {Padin},
  {Plagge}, {Pryke}, {Reichardt}, {Rest}, {Ruel}, {Ruhl}, {Schaffer}, {Shaw},
  {Shirokoff}, {Song}, {Spieler}, {Stalder}, {Staniszewski}, {Stark}, {Stubbs},
  {Vanderlinde}, {Vieira}, {Vikhlinin}, {Williamson}, {Yang}, \&
  {Zahn}}]{andersson10}
{Andersson}, K., {et~al.} 2010, ArXiv e-prints, 1006.3068

\bibitem[{{Andreon} \& {Ettori}(1999)}]{andreon99}
{Andreon}, S., \& {Ettori}, S. 1999, \apj, 516, 647

\bibitem[{{Barkhouse} {et~al.}(2007){Barkhouse}, {Yee}, \&
  {L{\'o}pez-Cruz}}]{barkhouse07}
{Barkhouse}, W.~A., {Yee}, H.~K.~C., \& {L{\'o}pez-Cruz}, O. 2007, \apj, 671,
  1471

\bibitem[{{Bartelmann}(1996)}]{Bartelmann96}
{Bartelmann}, M. 1996, \aap, 313, 697

\bibitem[{{Berlind} {et~al.}(2003){Berlind}, {Weinberg}, {Benson}, {Baugh},
  {Cole}, {Dav{\'e}}, {Frenk}, {Jenkins}, {Katz}, \& {Lacey}}]{berlind03}
{Berlind}, A.~A., {et~al.} 2003, \apj, 593, 1

\bibitem[{{Bertin}(2006)}]{bertin06}
{Bertin}, E. 2006, in Astronomical Society of the Pacific Conference Series,
  Vol. 351, Astronomical Data Analysis Software and Systems XV, ed.
  C.~{Gabriel}, C.~{Arviset}, D.~{Ponz}, \& S.~{Enrique}, 112--+

\bibitem[{{Bertin} \& {Arnouts}(1996)}]{bertin96}
{Bertin}, E., \& {Arnouts}, S. 1996, \aaps, 117, 393

\bibitem[{{Bertin} {et~al.}(2002){Bertin}, {Mellier}, {Radovich}, {Missonnier},
  {Didelon}, \& {Morin}}]{bertin02}
{Bertin}, E., {Mellier}, Y., {Radovich}, M., {Missonnier}, G., {Didelon}, P.,
  \& {Morin}, B. 2002, in Astronomical Society of the Pacific Conference
  Series, Vol. 281, Astronomical Data Analysis Software and Systems XI, ed.
  {D.~A.~Bohlender, D.~Durand, \& T.~H.~Handley}, 228--+

\bibitem[{Birkinshaw(1999)}]{birkinshaw99}
Birkinshaw, M. 1999, Physics Reports, 310, 97

\bibitem[{{Biviano} \& {Poggianti}(2009)}]{biviano09}
{Biviano}, A., \& {Poggianti}, B.~M. 2009, ArXiv e-prints, 0912.0365

\bibitem[{{B{\"o}hringer} {et~al.}(2004){B{\"o}hringer}, {Schuecker}, {Guzzo},
  {Collins}, {Voges}, {Cruddace}, {Ortiz-Gil}, {Chincarini}, {De Grandi},
  {Edge}, {MacGillivray}, {Neumann}, {Schindler}, \& {Shaver}}]{bohringer04}
{B{\"o}hringer}, H., {et~al.} 2004, \aap, 425, 367

\bibitem[{{Brimioulle} {et~al.}(2008){Brimioulle}, {Lerchster}, {Seitz},
  {Bender}, \& {Snigula}}]{brimioulle08}
{Brimioulle}, F., {Lerchster}, M., {Seitz}, S., {Bender}, R., \& {Snigula}, J.
  2008, ArXiv e-prints, 0811.3211

\bibitem[{{Brodwin} {et~al.}(2010){Brodwin}, {Ruel}, {Ade}, {Aird},
  {Andersson}, {Ashby}, {Bautz}, {Bazin}, {Benson}, {Bleem}, {Carlstrom},
  {Chang}, {Crawford}, {Crites}, {de Haan}, {Desai}, {Dobbs}, {Dudley},
  {Fazio}, {Foley}, {Forman}, {Garmire}, {George}, {Gladders}, {Gonzalez},
  {Halverson}, {High}, {Holder}, {Holzapfel}, {Hrubes}, {Jones}, {Joy},
  {Keisler}, {Knox}, {Lee}, {Leitch}, {Lueker}, {Marrone}, {McMahon}, {Mehl},
  {Meyer}, {Mohr}, {Montroy}, {Murray}, {Padin}, {Plagge}, {Pryke},
  {Reichardt}, {Rest}, {Ruhl}, {Schaffer}, {Shaw}, {Shirokoff}, {Song},
  {Spieler}, {Stalder}, {Stanford}, {Staniszewski}, {Stark}, {Stubbs},
  {Vanderlinde}, {Vieira}, {Vikhlinin}, {Williamson}, {Yang}, {Zahn}, \&
  {Zenteno}}]{brodwin10}
{Brodwin}, M., {et~al.} 2010, \apj, 721, 90

\bibitem[{{Bruzual} \& {Charlot}(2003)}]{bruzual03}
{Bruzual}, G., \& {Charlot}, S. 2003, \mnras, 344, 1000

\bibitem[{{Butcher} \& {Oemler}(1984)}]{butcher84}
{Butcher}, H., \& {Oemler}, Jr., A. 1984, \apj, 285, 426

\bibitem[{{Carlberg} {et~al.}(1996){Carlberg}, {Yee}, {Ellingson}, {Abraham},
  {Gravel}, {Morris}, \& {Pritchet}}]{carlberg96}
{Carlberg}, R.~G., {Yee}, H.~K.~C., {Ellingson}, E., {Abraham}, R., {Gravel},
  P., {Morris}, S., \& {Pritchet}, C.~J. 1996, \apj, 462, 32

\bibitem[{{Carlberg} {et~al.}(1997){Carlberg}, {Yee}, {Ellingson}, {Morris},
  {Abraham}, {Gravel}, {Pritchet}, {Smecker-Hane}, {Hartwick}, {Hesser},
  {Hutchings}, \& {Oke}}]{carlberg97}
{Carlberg}, R.~G., {et~al.} 1997, \apjl, 485, L13+

\bibitem[{{Carlstrom} {et~al.}(2009){Carlstrom}, {Ade}, {Aird}, {Benson},
  {Bleem}, {Busetti}, {Chang}, {Chauvin}, {Cho}, {Crawford}, {Crites}, {Dobbs},
  {Halverson}, {Heimsath}, {Holzapfel}, {Hrubes}, {Joy}, {Keisler}, {Lanting},
  {Lee}, {Leitch}, {Leong}, {Lu}, {Lueker}, {McMahon}, {Mehl}, {Meyer}, {Mohr},
  {Montroy}, {Padin}, {Plagge}, {Pryke}, {Ruhl}, {Schaffer}, {Schwan},
  {Shirokoff}, {Spieler}, {Staniszewski}, {Stark}, \& {Vieira}}]{carlstrom09}
{Carlstrom}, J.~E., {et~al.} 2009, submitted to \pasp, arXiv:0907.4445

\bibitem[{{Carlstrom} {et~al.}(2002){Carlstrom}, {Holder}, \&
  {Reese}}]{carlstrom02}
{Carlstrom}, J.~E., {Holder}, G.~P., \& {Reese}, E.~D. 2002, \araa, 40, 643

\bibitem[{{Cohn} {et~al.}(2007){Cohn}, {Evrard}, {White}, {Croton}, \&
  {Ellingson}}]{cohn07}
{Cohn}, J.~D., {Evrard}, A.~E., {White}, M., {Croton}, D., \& {Ellingson}, E.
  2007, \mnras, 382, 1738

\bibitem[{{Cole} {et~al.}(2000){Cole}, {Lacey}, {Baugh}, \& {Frenk}}]{cole00}
{Cole}, S., {Lacey}, C.~G., {Baugh}, C.~M., \& {Frenk}, C.~S. 2000, \mnras,
  319, 168

\bibitem[{{Collins} {et~al.}(1995){Collins}, {Guzzo}, {Nichol}, \&
  {Lumsden}}]{collins95}
{Collins}, C.~A., {Guzzo}, L., {Nichol}, R.~C., \& {Lumsden}, S.~L. 1995,
  \mnras, 274, 1071

\bibitem[{{De Lucia} {et~al.}(2004){De Lucia}, {Poggianti},
  {Arag{\'o}n-Salamanca}, {Clowe}, {Halliday}, {Jablonka}, {Milvang-Jensen},
  {Pell{\'o}}, {Poirier}, {Rudnick}, {Saglia}, {Simard}, \&
  {White}}]{delucia04}
{De Lucia}, G., {et~al.} 2004, \apjl, 610, L77

\bibitem[{{De Propris} {et~al.}(2003){De Propris}, {Colless}, {Driver},
  {Couch}, {Peacock}, {Baldry}, {Baugh}, {Bland-Hawthorn}, {Bridges}, {Cannon},
  {Cole}, {Collins}, {Cross}, {Dalton}, {Efstathiou}, {Ellis}, {Frenk},
  {Glazebrook}, {Hawkins}, {Jackson}, {Lahav}, {Lewis}, {Lumsden}, {Maddox},
  {Madgwick}, {Norberg}, {Percival}, {Peterson}, {Sutherland}, \&
  {Taylor}}]{depropris03}
{De Propris}, R., {et~al.} 2003, \mnras, 342, 725

\bibitem[{{De Propris} {et~al.}(2004){De Propris}, {Colless}, {Peacock},
  {Couch}, {Driver}, {Balogh}, {Baldry}, {Baugh}, {Bland-Hawthorn}, {Bridges},
  {Cannon}, {Cole}, {Collins}, {Cross}, {Dalton}, {Efstathiou}, {Ellis},
  {Frenk}, {Glazebrook}, {Hawkins}, {Jackson}, {Lahav}, {Lewis}, {Lumsden},
  {Maddox}, {Madgwick}, {Norberg}, {Percival}, {Peterson}, {Sutherland}, \&
  {Taylor}}]{depropris04}
------. 2004, \mnras, 351, 125

\bibitem[{{de Propris} {et~al.}(1999){de Propris}, {Stanford}, {Eisenhardt},
  {Dickinson}, \& {Elston}}]{depropris99}
{de Propris}, R., {Stanford}, S.~A., {Eisenhardt}, P.~R., {Dickinson}, M., \&
  {Elston}, R. 1999, \aj, 118, 719

\bibitem[{{De Propris} {et~al.}(2007){De Propris}, {Stanford}, {Eisenhardt},
  {Holden}, \& {Rosati}}]{depropris07}
{De Propris}, R., {Stanford}, S.~A., {Eisenhardt}, P.~R., {Holden}, B.~P., \&
  {Rosati}, P. 2007, \aj, 133, 2209

\bibitem[{{Ebeling} {et~al.}(1996){Ebeling}, {Voges}, {Bohringer}, {Edge},
  {Huchra}, \& {Briel}}]{ebeling96}
{Ebeling}, H., {Voges}, W., {Bohringer}, H., {Edge}, A.~C., {Huchra}, J.~P., \&
  {Briel}, U.~G. 1996, \mnras, 281, 799

\bibitem[{{Eisenhardt} {et~al.}(2008){Eisenhardt}, {Brodwin}, {Gonzalez},
  {Stanford}, {Stern}, {Barmby}, {Brown}, {Dawson}, {Dey}, {Doi}, {Galametz},
  {Jannuzi}, {Kochanek}, {Meyers}, {Morokuma}, \& {Moustakas}}]{eisenhardt08}
{Eisenhardt}, P.~R.~M., {et~al.} 2008, \apj, 684, 905

\bibitem[{{Elston} {et~al.}(2006){Elston}, {Gonzalez}, {McKenzie}, {Brodwin},
  {Brown}, {Cardona}, {Dey}, {Dickinson}, {Eisenhardt}, {Jannuzi}, {Lin},
  {Mohr}, {Raines}, {Stanford}, \& {Stern}}]{elston06}
{Elston}, R.~J., {et~al.} 2006, \apj, 639, 816

\bibitem[{{Fairley} {et~al.}(2002){Fairley}, {Jones}, {Wake}, {Collins},
  {Burke}, {Nichol}, \& {Romer}}]{fairley02}
{Fairley}, B.~W., {Jones}, L.~R., {Wake}, D.~A., {Collins}, C.~A., {Burke},
  D.~J., {Nichol}, R.~C., \& {Romer}, A.~K. 2002, \mnras, 330, 755

\bibitem[{{Gehrels}(1986)}]{gehrels86}
{Gehrels}, N. 1986, \apj, 303, 336

\bibitem[{{Giacconi} {et~al.}(1972){Giacconi}, {Murray}, {Gursky}, {Kellogg},
  {Schreier}, \& {Tananbaum}}]{giacconi72}
{Giacconi}, R., {Murray}, S., {Gursky}, H., {Kellogg}, E., {Schreier}, E., \&
  {Tananbaum}, H. 1972, \apj, 178, 281

\bibitem[{{Gladders} {et~al.}(1998){Gladders}, {Lopez-Cruz}, {Yee}, \&
  {Kodama}}]{gladders98}
{Gladders}, M.~D., {Lopez-Cruz}, O., {Yee}, H.~K.~C., \& {Kodama}, T. 1998,
  \apj, 501, 571

\bibitem[{{Goto} {et~al.}(2004){Goto}, {Yagi}, {Tanaka}, \& {Okamura}}]{goto04}
{Goto}, T., {Yagi}, M., {Tanaka}, M., \& {Okamura}, S. 2004, \mnras, 348, 515

\bibitem[{{Guzzo} {et~al.}(1999){Guzzo}, {B{\"o}hringer}, {Schuecker},
  {Collins}, {Schindler}, {Neumann}, {de Grandi}, {Cruddace}, {Chincarini},
  {Edge}, {Shaver}, \& {Voges}}]{guzzo99}
{Guzzo}, L., {et~al.} 1999, The Messenger, 95, 27

\bibitem[{{Hansen} {et~al.}(2009){Hansen}, {Sheldon}, {Wechsler}, \&
  {Koester}}]{hansen09}
{Hansen}, S.~M., {Sheldon}, E.~S., {Wechsler}, R.~H., \& {Koester}, B.~P. 2009,
  \apj, 699, 1333

\bibitem[{{High} {et~al.}(2010){High}, {Stalder}, {Song}, {Ade}, {Aird},
  {Allam}, {Armstrong}, {Barkhouse}, {Benson}, {Bertin}, {Bhattacharya},
  {Bleem}, {Brodwin}, {Buckley-Geer}, {Carlstrom}, {Challis}, {Chang},
  {Crawford}, {Crites}, {de Haan}, {Desai}, {Dobbs}, {Dudley}, {Foley},
  {George}, {Gladders}, {Halverson}, {Hamuy}, {Hansen}, {Holder}, {Holzapfel},
  {Hrubes}, {Joy}, {Keisler}, {Lee}, {Leitch}, {Lin}, {Lin}, {Loehr}, {Lueker},
  {Marrone}, {McMahon}, {Mehl}, {Meyer}, {Mohr}, {Montroy}, {Morell}, {Ngeow},
  {Padin}, {Plagge}, {Pryke}, {Reichardt}, {Rest}, {Ruel}, {Ruhl}, {Schaffer},
  {Shaw}, {Shirokoff}, {Smith}, {Spieler}, {Staniszewski}, {Stark}, {Stubbs},
  {Tucker}, {Vanderlinde}, {Vieira}, {Williamson}, {Wood-Vasey}, {Yang},
  {Zahn}, \& {Zenteno}}]{high10}
{High}, F.~W., {et~al.} 2010, \apj, 723, 1736

\bibitem[{{High} {et~al.}(2009){High}, {Stubbs}, {Rest}, {Stalder}, \&
  {Challis}}]{high09}
{High}, F.~W., {Stubbs}, C.~W., {Rest}, A., {Stalder}, B., \& {Challis}, P.
  2009, \aj, 138, 110

\bibitem[{{Holden} {et~al.}(2004){Holden}, {Stanford}, {Eisenhardt}, \&
  {Dickinson}}]{holden04}
{Holden}, B.~P., {Stanford}, S.~A., {Eisenhardt}, P., \& {Dickinson}, M. 2004,
  \aj, 127, 2484

\bibitem[{{Johnston} {et~al.}(2007){Johnston}, {Sheldon}, {Wechsler}, {Rozo},
  {Koester}, {Frieman}, {McKay}, {Evrard}, {Becker}, \& {Annis}}]{johnston07}
{Johnston}, D.~E., {et~al.} 2007, ArXiv e-prints, 0709.1159

\bibitem[{{Kodama} \& {Bower}(2001)}]{kodama01}
{Kodama}, T., \& {Bower}, R.~G. 2001, \mnras, 321, 18

\bibitem[{{Koester} {et~al.}(2007){Koester}, {McKay}, {Annis}, {Wechsler},
  {Evrard}, {Bleem}, {Becker}, {Johnston}, {Sheldon}, {Nichol}, {Miller},
  {Scranton}, {Bahcall}, {Barentine}, {Brewington}, {Brinkmann}, {Harvanek},
  {Kleinman}, {Krzesinski}, {Long}, {Nitta}, {Schneider}, {Sneddin}, {Voges},
  \& {York}}]{koester07a}
{Koester}, B.~P., {et~al.} 2007, \apj, 660, 239

\bibitem[{{Komatsu} {et~al.}(2010){Komatsu}, {Smith}, {Dunkley}, {Bennett},
  {Gold}, {Hinshaw}, {Jarosik}, {Larson}, {Nolta}, {Page}, {Spergel},
  {Halpern}, {Hill}, {Kogut}, {Limon}, {Meyer}, {Odegard}, {Tucker}, {Weiland},
  {Wollack}, \& {Wright}}]{komatsu10}
{Komatsu}, E., {et~al.} 2010, accepted by ApJS, arXiv:1001.4538

\bibitem[{{Lin} {et~al.}(2006){Lin}, {Mohr}, {Gonzalez}, \& {Stanford}}]{lin06}
{Lin}, Y., {Mohr}, J.~J., {Gonzalez}, A.~H., \& {Stanford}, S.~A. 2006, \apjl,
  650, L99

\bibitem[{{Lin} {et~al.}(2003){Lin}, {Mohr}, \& {Stanford}}]{lin03b}
{Lin}, Y., {Mohr}, J.~J., \& {Stanford}, S.~A. 2003, \apj, 591, 749

\bibitem[{{Lin} {et~al.}(2004){Lin}, {Mohr}, \& {Stanford}}]{lin04a}
------. 2004, \apj, 610, 745

\bibitem[{{Loh} {et~al.}(2008){Loh}, {Ellingson}, {Yee}, {Gilbank}, {Gladders},
  \& {Barrientos}}]{loh08}
{Loh}, Y., {Ellingson}, E., {Yee}, H.~K.~C., {Gilbank}, D.~G., {Gladders},
  M.~D., \& {Barrientos}, L.~F. 2008, \apj, 680, 214

\bibitem[{{L{\'o}pez-Cruz} {et~al.}(2004){L{\'o}pez-Cruz}, {Barkhouse}, \&
  {Yee}}]{lopez04}
{L{\'o}pez-Cruz}, O., {Barkhouse}, W.~A., \& {Yee}, H.~K.~C. 2004, \apj, 614,
  679

\bibitem[{{Lucey}(1983)}]{lucey83}
{Lucey}, J.~R. 1983, \mnras, 204, 33

\bibitem[{{Margoniner} \& {de Carvalho}(2000)}]{margoniner00}
{Margoniner}, V.~E., \& {de Carvalho}, R.~R. 2000, \aj, 119, 1562

\bibitem[{{Marinoni} \& {Hudson}(2002)}]{marinoni02}
{Marinoni}, C., \& {Hudson}, M.~J. 2002, \apj, 569, 101

\bibitem[{{McInnes} {et~al.}(2009){McInnes}, {Menanteau}, {Heavens}, {Hughes},
  {Jimenez}, {Massey}, {Simon}, \& {Taylor}}]{mcinnes09}
{McInnes}, R.~N., {Menanteau}, F., {Heavens}, A.~F., {Hughes}, J.~P.,
  {Jimenez}, R., {Massey}, R., {Simon}, P., \& {Taylor}, A. 2009, \mnras, 399,
  L84

\bibitem[{{Menanteau} {et~al.}(2009){Menanteau}, {Hughes}, {Jimenez},
  {Hernandez-Monteagudo}, {Verde}, {Kosowsky}, {Moodley}, {Infante}, \&
  {Roche}}]{menanteau09}
{Menanteau}, F., {et~al.} 2009, \apj, 698, 1221

\bibitem[{{Mohr} {et~al.}(2008){Mohr}, {Adams}, {Barkhouse}, {Beldica},
  {Bertin}, {Cai}, {da Costa}, {Darnell}, {Daues}, {Jarvis}, {Gower}, {Lin},
  {Martelli}, {Neilsen}, {Ngeow}, {Ogando}, {Parga}, {Sheldon}, {Tucker},
  {Kuropatkin}, \& {Stoughton}}]{mohr08}
{Mohr}, J.~J., {et~al.} 2008, in Society of Photo-Optical Instrumentation
  Engineers (SPIE) Conference Series, Vol. 7016, Society of Photo-Optical
  Instrumentation Engineers (SPIE) Conference Series

\bibitem[{{Muzzin} {et~al.}(2008){Muzzin}, {Wilson}, {Lacy}, {Yee}, \&
  {Stanford}}]{muzzin08}
{Muzzin}, A., {Wilson}, G., {Lacy}, M., {Yee}, H.~K.~C., \& {Stanford}, S.~A.
  2008, \apj, 686, 966

\bibitem[{{Muzzin} {et~al.}(2007{\natexlab{a}}){Muzzin}, {Yee}, {Hall},
  {Ellingson}, \& {Lin}}]{muzzin07a}
{Muzzin}, A., {Yee}, H.~K.~C., {Hall}, P.~B., {Ellingson}, E., \& {Lin}, H.
  2007{\natexlab{a}}, \apj, 659, 1106

\bibitem[{{Muzzin} {et~al.}(2007{\natexlab{b}}){Muzzin}, {Yee}, {Hall}, \&
  {Lin}}]{muzzin07b}
{Muzzin}, A., {Yee}, H.~K.~C., {Hall}, P.~B., \& {Lin}, H. 2007{\natexlab{b}},
  \apj, 663, 150

\bibitem[{{Muzzin} {et~al.}(2007{\natexlab{c}}){Muzzin}, {Yee}, {Hall}, \&
  {Lin}}]{muzzin07}
------. 2007{\natexlab{c}}, \apj, 663, 150

\bibitem[{{Nagai} \& {Kravtsov}(2005)}]{nagai05}
{Nagai}, D., \& {Kravtsov}, A.~V. 2005, \apj, 618, 557

\bibitem[{{Navarro} {et~al.}(1997){Navarro}, {Frenk}, \& {White}}]{navarro97}
{Navarro}, J.~F., {Frenk}, C.~S., \& {White}, S.~D.~M. 1997, \apj, 490, 493

\bibitem[{{Newberry} {et~al.}(1988){Newberry}, {Kirshner}, \&
  {Boroson}}]{newberry88}
{Newberry}, M.~V., {Kirshner}, R.~P., \& {Boroson}, T.~A. 1988, \apj, 335, 629

\bibitem[{{Ngeow} {et~al.}(2006){Ngeow}, {Mohr}, {Alam}, {Barkhouse},
  {Beldica}, {Cai}, {Daues}, {Plante}, {Annis}, {Lin}, {Tucker}, \&
  {Smith}}]{ngeow06}
{Ngeow}, C., {et~al.} 2006, in Society of Photo-Optical Instrumentation
  Engineers (SPIE) Conference Series, Vol. 6270, Society of Photo-Optical
  Instrumentation Engineers (SPIE) Conference Series

\bibitem[{{Paolillo} {et~al.}(2001){Paolillo}, {Andreon}, {Longo}, {Puddu},
  {Gal}, {Scaramella}, {Djorgovski}, \& {de Carvalho}}]{paolillo01}
{Paolillo}, M., {Andreon}, S., {Longo}, G., {Puddu}, E., {Gal}, R.~R.,
  {Scaramella}, R., {Djorgovski}, S.~G., \& {de Carvalho}, R. 2001, \aap, 367,
  59

\bibitem[{{Popesso} {et~al.}(2007{\natexlab{a}}){Popesso}, {Biviano},
  {B{\"o}hringer}, \& {Romaniello}}]{popesso07b}
{Popesso}, P., {Biviano}, A., {B{\"o}hringer}, H., \& {Romaniello}, M.
  2007{\natexlab{a}}, \aap, 464, 451

\bibitem[{{Popesso} {et~al.}(2007{\natexlab{b}}){Popesso}, {Biviano},
  {Romaniello}, \& {B{\"o}hringer}}]{popesso07}
{Popesso}, P., {Biviano}, A., {Romaniello}, M., \& {B{\"o}hringer}, H.
  2007{\natexlab{b}}, \aap, 461, 411

\bibitem[{{Popesso} {et~al.}(2005){Popesso}, {B{\"o}hringer}, {Romaniello}, \&
  {Voges}}]{popessoII05}
{Popesso}, P., {B{\"o}hringer}, H., {Romaniello}, M., \& {Voges}, W. 2005,
  \aap, 433, 415

\bibitem[{Press \& Schechter(1974)}]{press74}
Press, W., \& Schechter, P. 1974, \apj, 187, 425

\bibitem[{{Press} {et~al.}(1992){Press}, {Teukolsky}, {Vetterling}, \&
  {Flannery}}]{press92}
{Press}, W.~H., {Teukolsky}, S.~A., {Vetterling}, W.~T., \& {Flannery}, B.~P.
  1992, {Numerical recipes in C. The art of scientific computing}, ed. {Press,
  W.~H., Teukolsky, S.~A., Vetterling, W.~T., \& Flannery, B.~P. } (Cambridge:
  University Press, |c1992, 2nd ed.)

\bibitem[{{Rakos} \& {Schombert}(1995)}]{rakos95}
{Rakos}, K.~D., \& {Schombert}, J.~M. 1995, \apj, 439, 47

\bibitem[{{Rines} {et~al.}(2004){Rines}, {Geller}, {Diaferio}, {Kurtz}, \&
  {Jarrett}}]{rines04}
{Rines}, K., {Geller}, M.~J., {Diaferio}, A., {Kurtz}, M.~J., \& {Jarrett},
  T.~H. 2004, \aj, 128, 1078

\bibitem[{{Romeo} {et~al.}(2005){Romeo}, {Portinari}, \&
  {Sommer-Larsen}}]{romeo05}
{Romeo}, A.~D., {Portinari}, L., \& {Sommer-Larsen}, J. 2005, \mnras, 361, 983

\bibitem[{{Roncarelli} {et~al.}(2010){Roncarelli}, {Pointecouteau}, {Giard},
  {Montier}, \& {Pello}}]{roncarelli10}
{Roncarelli}, M., {Pointecouteau}, E., {Giard}, M., {Montier}, L., \& {Pello},
  R. 2010, \aap, 512, A20+

\bibitem[{{Saro} {et~al.}(2006){Saro}, {Borgani}, {Tornatore}, {Dolag},
  {Murante}, {Biviano}, {Calura}, \& {Charlot}}]{saro06}
{Saro}, A., {Borgani}, S., {Tornatore}, L., {Dolag}, K., {Murante}, G.,
  {Biviano}, A., {Calura}, F., \& {Charlot}, S. 2006, \mnras, 373, 397

\bibitem[{{Schechter}(1976)}]{schechter76}
{Schechter}, P. 1976, \apj, 203, 297

\bibitem[{{Seljak}(2000)}]{seljak00a}
{Seljak}, U. 2000, \mnras, 318, 203

\bibitem[{{Smail} {et~al.}(1998){Smail}, {Edge}, {Ellis}, \&
  {Blandford}}]{smail98}
{Smail}, I., {Edge}, A.~C., {Ellis}, R.~S., \& {Blandford}, R.~D. 1998, \mnras,
  293, 124

\bibitem[{{Stanford} {et~al.}(2005){Stanford}, {Eisenhardt}, {Brodwin},
  {Gonzalez}, {Stern}, {Jannuzi}, {Dey}, {Brown}, {McKenzie}, \&
  {Elston}}]{stanford05}
{Stanford}, S.~A., {et~al.} 2005, \apjl, 634, L129

\bibitem[{{Stanford} {et~al.}(1997){Stanford}, {Elston}, {Eisenhardt},
  {Spinrad}, {Stern}, \& {Dey}}]{stanford97}
{Stanford}, S.~A., {Elston}, R., {Eisenhardt}, P.~R., {Spinrad}, H., {Stern},
  D., \& {Dey}, A. 1997, \aj, 114, 2232

\bibitem[{{Staniszewski} {et~al.}(2009){Staniszewski}, {Ade}, {Aird}, {Benson},
  {Bleem}, {Carlstrom}, {Chang}, {Cho}, {Crawford}, {Crites}, {de Haan},
  {Dobbs}, {Halverson}, {Holder}, {Holzapfel}, {Hrubes}, {Joy}, {Keisler},
  {Lanting}, {Lee}, {Leitch}, {Loehr}, {Lueker}, {McMahon}, {Mehl}, {Meyer},
  {Mohr}, {Montroy}, {Ngeow}, {Padin}, {Plagge}, {Pryke}, {Reichardt}, {Ruhl},
  {Schaffer}, {Shaw}, {Shirokoff}, {Spieler}, {Stalder}, {Stark},
  {Vanderlinde}, {Vieira}, {Zahn}, \& {Zenteno}}]{staniszewski09}
{Staniszewski}, Z., {et~al.} 2009, \apj, 701, 32

\bibitem[{{Sunyaev} \& {Zel'dovich}(1972)}]{sunyaev72}
{Sunyaev}, R.~A., \& {Zel'dovich}, Y.~B. 1972, Comments on Astrophysics and
  Space Physics, 4, 173

\bibitem[{{Sutherland}(1988)}]{sutherland88}
{Sutherland}, W. 1988, \mnras, 234, 159

\bibitem[{{Szalay} {et~al.}(1999){Szalay}, {Connolly}, \& {Szokoly}}]{szalay99}
{Szalay}, A.~S., {Connolly}, A.~J., \& {Szokoly}, G.~P. 1999, \aj, 117, 68

\bibitem[{{Toft} {et~al.}(2004){Toft}, {Mainieri}, {Rosati}, {Lidman},
  {Demarco}, {Nonino}, \& {Stanford}}]{toft04}
{Toft}, S., {Mainieri}, V., {Rosati}, P., {Lidman}, C., {Demarco}, R.,
  {Nonino}, M., \& {Stanford}, S.~A. 2004, \aap, 422, 29

\bibitem[{{Urquhart} {et~al.}(2010){Urquhart}, {Willis}, {Hoekstra}, \&
  {Pierre}}]{urquhart10}
{Urquhart}, S.~A., {Willis}, J.~P., {Hoekstra}, H., \& {Pierre}, M. 2010, ArXiv
  e-prints, 1004.0020

\bibitem[{{van Breukelen} {et~al.}(2006){van Breukelen}, {Clewley}, {Bonfield},
  {Rawlings}, {Jarvis}, {Barr}, {Foucaud}, {Almaini}, {Cirasuolo}, {Dalton},
  {Dunlop}, {Edge}, {Hirst}, {McLure}, {Page}, {Sekiguchi}, {Simpson}, {Smail},
  \& {Watson}}]{vanbreukelen06}
{van Breukelen}, C., {et~al.} 2006, \mnras, 373, L26

\bibitem[{{Vanderlinde} {et~al.}(2010){Vanderlinde}, {Crawford}, {de Haan},
  {Dudley}, {Shaw}, {Ade}, {Aird}, {Benson}, {Bleem}, {Brodwin}, {Carlstrom},
  {Chang}, {Crites}, {Desai}, {Dobbs}, {Foley}, {George}, {Gladders}, {Hall},
  {Halverson}, {High}, {Holder}, {Holzapfel}, {Hrubes}, {Joy}, {Keisler},
  {Knox}, {Lee}, {Leitch}, {Loehr}, {Lueker}, {Marrone}, {McMahon}, {Mehl},
  {Meyer}, {Mohr}, {Montroy}, {Ngeow}, {Padin}, {Plagge}, {Pryke}, {Reichardt},
  {Rest}, {Ruel}, {Ruhl}, {Schaffer}, {Shirokoff}, {Song}, {Spieler},
  {Stalder}, {Staniszewski}, {Stark}, {Stubbs}, {van Engelen}, {Vieira},
  {Williamson}, {Yang}, {Zahn}, \& {Zenteno}}]{vanderlinde10}
{Vanderlinde}, K., {et~al.} 2010, submitted to \apj, arXiv:1003.0003

\bibitem[{{Vikhlinin} {et~al.}(2009){Vikhlinin}, {Kravtsov}, {Burenin},
  {Ebeling}, {Forman}, {Hornstrup}, {Jones}, {Murray}, {Nagai}, {Quintana}, \&
  {Voevodkin}}]{vikhlinin09}
{Vikhlinin}, A., {et~al.} 2009, \apj, 692, 1060

\bibitem[{Vikhlinin {et~al.}(1998)Vikhlinin, McNamara, Forman, Jones, Quintana,
  \& Hornstrup}]{Vikhlinin98}
Vikhlinin, A., McNamara, B., Forman, W., Jones, C., Quintana, H., \& Hornstrup,
  A. 1998, \apj, 502, 558

\bibitem[{{Voges} {et~al.}(1999){Voges}, {Aschenbach}, {Boller},
  {Br{\"a}uninger}, {Briel}, {Burkert}, {Dennerl}, {Englhauser}, {Gruber},
  {Haberl}, {Hartner}, {Hasinger}, {K{\"u}rster}, {Pfeffermann}, {Pietsch},
  {Predehl}, {Rosso}, {Schmitt}, {Tr{\"u}mper}, \& {Zimmermann}}]{voges99}
{Voges}, W., {et~al.} 1999, \aap, 349, 389

\bibitem[{{Voges} {et~al.}(2000){Voges}, {Aschenbach}, {Boller}, {Brauninger},
  {Briel}, {Burkert}, {Dennerl}, {Englhauser}, {Gruber}, {Haberl}, {Hartner},
  {Hasinger}, {Pfeffermann}, {Pietsch}, {Predehl}, {Schmitt}, {Trumper}, \&
  {Zimmermann}}]{voges00}
------. 2000, VizieR Online Data Catalog, 9029, 0

\bibitem[{{Williamson} {et~al.}(2011){Williamson}, {Benson}, {High},
  {Vanderlinde}, {Ade}, {Aird}, {Andersson}, {Armstrong}, {Ashby}, {Bautz},
  {Bazin}, {Bertin}, {Bleem}, {Bonamente}, {Brodwin}, {Carlstrom}, {Chang},
  {Clocchiatti}, {Crawford}, {Crites}, {de Haan}, {Desai}, {Dobbs}, {Dudley},
  {Fazio}, {Foley}, {Forman}, {Garmire}, {George}, {Gladders}, {Gonzalez},
  {Halverson}, {Holder}, {Holzapfel}, {Hoover}, {Hrubes}, {Jones}, {Joy},
  {Keisler}, {Knox}, {Lee}, {Leitch}, {Lueker}, {Luong-Van}, {Marrone},
  {McMahon}, {Mehl}, {Meyer}, {Mohr}, {Montroy}, {Murray}, {Padin}, {Plagge},
  {Pryke}, {Reichardt}, {Rest}, {Ruel}, {Ruhl}, {Saliwanchik}, {Saro},
  {Schaffer}, {Shaw}, {Shirokoff}, {Song}, {Spieler}, {Stalder}, {Stanford},
  {Staniszewski}, {Stark}, {Story}, {Stubbs}, {Vieira}, {Vikhlinin}, \&
  {Zenteno}}]{williamson11}
{Williamson}, R., {et~al.} 2011, ArXiv e-prints, 1101.1290

\bibitem[{{Zhang} {et~al.}(2006){Zhang}, {B{\"o}hringer}, {Finoguenov},
  {Ikebe}, {Matsushita}, {Schuecker}, {Guzzo}, \& {Collins}}]{zhang06}
{Zhang}, Y.-Y., {B{\"o}hringer}, H., {Finoguenov}, A., {Ikebe}, Y.,
  {Matsushita}, K., {Schuecker}, P., {Guzzo}, L., \& {Collins}, C.~A. 2006,
  \aap, 456, 55

\bibitem[{{Zheng} \& {Weinberg}(2007)}]{zheng07}
{Zheng}, Z., \& {Weinberg}, D.~H. 2007, \apj, 659, 1

\end{thebibliography}
\end{document}